\documentclass[twocolumn]{aastex62}

\newcommand{\HI}{\text{H\,{\sc i}}}

\newcommand{\taueff}{\tau_\mathrm{eff}}

\usepackage{subfigure}
\usepackage{comment}
\usepackage{amsmath}

\begin{document}
\title{Evidence for a highly opaque large-scale galaxy void at the end of reionization}

\author[0000-0001-9044-1747]{Daichi Kashino}
\affiliation{Department of Physics, ETH Z{\" u}rich, Wolfgang-Pauli-Strasse 27, CH-8093 Z{\" u}rich, Switzerland}
\author[0000-0002-6423-3597]{Simon J.~Lilly}
\affiliation{Department of Physics, ETH Z{\" u}rich, Wolfgang-Pauli-Strasse 27, CH-8093 Z{\" u}rich, Switzerland}
\author{Takatoshi Shibuya}
\affiliation{Kitami Institute of Technology, 165 Koen-cho, Kitami, Hokkaido 090-8507, Japan}
\author[0000-0002-1049-6658]{Masami Ouchi}
\affiliation{Division of Science, National Astronomical Observatory of Japan, 2-21-1 Osawa, Mitaka, Tokyo 181-8588, Japan}
\affiliation{Institute for Cosmic Ray Research, The University of Tokyo, 5-1-5 Kashiwanoha, Kashiwa, Chiba 277-8583, Japan}
\author[0000-0001-5493-6259]{Nobunari Kashikawa}
\affiliation{Department of Astronomy, Graduate School of Science, The University of Tokyo, 7-3-1 Hongo, Bunkyo-ku, Tokyo 113-0033, Japan}

\correspondingauthor{Daichi Kashino}
\email{kashinod@phys.ethz.ch}

\begin{abstract}

We present evidence that a region of high effective Ly$\alpha$ optical depth at $z\sim5.7$ is associated with an underdense region at the tail end of cosmic reionization.  We carried out a survey of Lyman-break Galaxies (LBGs) using Subaru Hyper Suprime-Cam in the field of the $z=5.98$ quasar J0148+0600, whose spectrum presents an unusually long ($\sim 160 ~ \mathrm{cMpc}$) and opaque ($\tau \gtrsim7$) Ly$\alpha$ trough at $5.5\le z\le 5.9$.  LBG candidates were selected to lie within the redshift range of the trough, and the projected number densities were measured within 90~cMpc of the quasar sightline.  The region within $8\arcmin$ (or $\approx 19~\mathrm{cMpc}$) of the quasar position is the most underdense of the whole field.  The significance of the presence of the void is estimated to be $99\%$.  This is consistent with the significant deficit of Ly$\alpha$ emitters (LAEs) at $z=5.72$ reported by Becker et al. and suggests that the paucity of LAEs is not purely due to the removal of the Ly$\alpha$ emission by the high opacity but reflects a real coherent underdensity of galaxies across the entire redshift range of the trough.  These observations are consistent with scenarios in which large optical depth fluctuations arise due to fluctuations in the galaxy-dominant UV background or due to residual neutral islands that are expected from reionization that is completed at redshifts as low as $z\lesssim5.5$.

\end{abstract}

\section{Introduction}

The evolution of the intergalactic medium (IGM) during the Epoch of Reionization (EoR) has long been of particular interest in astrophysics.  Measurements of the Gunn--Peterson troughs \citep{1965ApJ...142.1633G} in the Ly$\alpha$ forest spectra of distant quasars have established the increase in the average {\HI} opacity back to $z\approx6$, marking the end of reionization \citep{2002AJ....123.1247F,2006AJ....132..117F}.   At $z<5$, the probability distribution of the effective Ly$\alpha$ optical depth, $\taueff$, measured on scales of $\sim70~\mathrm{cMpc}$, can be explained by fluctuations in the density field, illuminated with a homogeneous ultraviolet background (UVB).  Interestingly, however, there are indications that the paradigm must change nearer to the EoR.  Deep quasar spectra have shown that the variations in $\taueff$ increase dramatically at $z>5.5$ to a level about three times that expected from density fluctuations alone \citep{2015MNRAS.447.3402B,2018MNRAS.479.1055B,2018ApJ...864...53E}.  In parallel, a large-scale spatial coherence is found, epitomized by the discovery of an enormously long ($\sim 160~\mathrm{cMpc}$) and opaque ($\taueff>7$) Ly$\alpha$ trough covering $z=5.52\text{--}5.88$ toward the quasar J0148$+$0600 at $z=5.98$ \citep{2015MNRAS.447.3402B}.

The presence of highly opaque regions and the surprising degree of spatial coherence, despite the universe as a whole being largely reionized by then, have motivated simulators to reproduce these observations by invoking additional fluctuations due to inhomogeneities in either the UVB or IGM temperature.  \citet{2016MNRAS.460.1328D} showed that a galaxy-dominated UVB with spatially varying mean free paths, $\lambda_\mathrm{mfp}$, can explain the observed $\taueff$ distribution (we refer to this scenario as the fluctuating-$\lambda_\mathrm{mfp}$ model), though the model requires a short average mean free path of $\left< \lambda_\mathrm{mfp} \right>\approx15~\mathrm{cMpc}$, i.e., a factor of $\gtrsim3$ shorter than measurements at $z\gtrsim5$.  However, \citet{2018MNRAS.473..560D} showed that the direct measurements are likely biased toward higher values and that the short $\lambda_\mathrm{mfp}$ values cannot be ruled out.   

\citet{2015MNRAS.453.2943C} proposed a model in which the large fluctuations and spatial coherence in the optical depth arise due to enhanced UVB fluctuations with significant contribution of rare, bright sources such as quasars (rare-source model; see also \citealt{2017MNRAS.465.3429C}).  This model, however, requires the number density of such quasars to be higher than that inferred from observations \citep[e.g.,][]{2018ApJ...869..150M}, and the required large contribution of quasars could lead to an earlier He\,{\sc ii} reionization than that suggested by observations \citep{2017MNRAS.468.4691D}.  

On the other hand, \citet{2015ApJ...813L..38D} present a scenario in which spatial variations in reionization redshifts through an extended history could cause relic fluctuations in the IGM temperature in the post-reionization era.  The temperature variations then affect the hydrogen recombination rate and thus enhance the $\taueff$ fluctuations (fluctuating-$T_\mathrm{IGM}$ model).

More recently, \citet{2019MNRAS.485L..24K} proposed another model in which reionization ended relatively late ($z\sim5.2$), having residual islands of neutral hydrogen until $z\le 5.5$ (late-reionization model).  With a slightly modified setup from that used in \citet{2019MNRAS.485L..24K}, \citet{2020MNRAS.491.1736K} showed that a trough as long as that seen in the J0148$+$0600 spectrum could be reproduced in underdense regions.  This scenario is different from the fluctuating-$\lambda_\mathrm{mfp}$ model where the global reionization indeed ended before the time we observe the troughs and the underdense regions are opaque for a relatively high, but still $\ll1$ neutral fraction $x_\text{H\,{\sc i}}$ due to a weaker UVB.  In the late-reionization model, the underdense regions are highly opaque because of the residual neutral islands where $x_\text{H\,{\sc i}}\sim1$.  These same underdense regions would then have low $\taueff$, once the ionizing radiation penetrates throughout the void, because the mean free path becomes long and the fluctuations in the UVB are quickly reduced.

These different scenarios can in principle be tested by observations.  Correlating the number density of tracers of the underlying density field (i.e., galaxies) with the H\,{\sc i} optical depth is a powerful way to distinguish the models \citep{2018ApJ...860..155D}.  Recently, \citet{2018ApJ...863...92B} reported that the long Ly$\alpha$ absorption trough toward J0148$+$0600 is associated with a significant deficit of Ly$\alpha$ emitters (LAEs) at $z=5.68\textrm{--}5.78$ ($\Delta z\approx0.1$).  Taken at face value, the result is consistent with the fluctuating-$\lambda_\mathrm{mfp}$ model and the late-reionization model.  However, as noted by the authors, it is not entirely sure whether LAEs trace the underlying density field.  In addition to a possible bias of LAEs relative to the overall galaxy population, a further concern is that the Ly$\alpha$ emission may itself be destroyed by neutral hydrogen in the high-$\taueff$ IGM around the galaxies and thus the detection of LAEs would be suppressed in these regions \citep{2017ApJ...839...44S,2020MNRAS.491.1736K}.  This could result in an apparent but misleading inverse correlation between $\taueff$ and the observed LAE surface density.  Moreover, the redshift coverage of LAEs identified by a narrowband filter ({\it NB}\,816; $\Delta z\approx0.1$) is not sufficient to explore the origin of the surprising spatial coherence over $\sim1 60~\mathrm{cMpc}$ ($\Delta z\approx0.4$) of the trough.  

In the first place, observations in more quasar fields, including those presenting low $\taueff$ ($\sim2$), are essential to establish the correlation between the optical depth and density in its entirety.  In particular, the variation in the galaxy density in low-$\taueff$ regions is key to distinguishing between the fluctuating-$\lambda_\mathrm{mpf}$ and late-reionization models.  In parallel, searches for rare, bright sources, especially near low-$\taueff$ regions, are required to test their possible contribution to the UVB fluctuations.  

To overcome these limitations and obtain a more conclusive statement, we are carrying out surveys of Lyman-break Galaxies (LBGs) across $5.5\lesssim z\lesssim 6.0$ using Subaru Hyper Suprime-Cam (HSC) in multiple quasar fields, including those with both extremely high and low $\taueff$ values.  LBG surveys are complementary to LAE surveys as different types of galaxies are used as density tracers and have the advantages that they are not affected by the IGM opacity and can also cover a wider redshift range.  

In this paper, we present our first result in the field of J0148$+$0600.  The layout of the paper is as follows.  
We present our HSC data in Section \ref{sec:data}.  
In Section \ref{sec:analysis}, we describe the selection of LBG candidates and present the results.  
We compare the results with models and give discussions in Section \ref{sec:discussions}.  
Finally, we summarize our results in Section \ref{sec:summary}.
Throughout the paper, we adopt a standard flat cosmology with $(\Omega_\mathrm{m},~\Omega_\Lambda,~h) = (0.308, 0.692, 0.678)$ \citep{2016A&A...594A..13P}, and all magnitudes are presented in the AB system.

\section{Data} \label{sec:data}

We acquired new deep HSC $z$-band imaging of a field centered on the quasar J0148$+$0600 ($z_\mathrm{em}=5.98$, $\textrm{R.A.}=01^\mathrm{h}48^\mathrm{m}37.\!^\mathrm{s}639$, $\textrm{decl.}=+06\arcdeg00\arcmin20.\!\arcsec01$) in 2018 November.  We incorporated additional public HSC data in the $r2$, $i2$, and {\it NB}\,816 bands that were taken in 2016 September and 2017 August.  These retrieved data are the set of images used for the LAE survey in \citet{2018ApJ...863...92B}.  The combined HSC data are summarized in Table \ref{tb:observations}.

\begin{deluxetable}{lccccc}
\tablecaption{Summary of Subaru/HSC data \label{tb:observations}}
\tablehead{
 		\colhead{Field}&
		\colhead{Filter}&
		\colhead{$t_\mathrm{exp}$}&
		\colhead{Seeing\tablenotemark{a} }&
		\colhead{$m_{5\sigma}\tablenotemark{b}$} \\  		
		\colhead{}&
		\colhead{}&
		\colhead{(hr)}&
		\colhead{(arcsec)}&
		\colhead{(mag)} }
\startdata
J0148$+$0600 & $r2$\tablenotemark{c} & 1.44 & 0.61 & 26.82 \\
& $i2$\tablenotemark{c} & 2.15 & 0.63 & 26.34\\
& $z$  & 1.33 & 0.78 & 25.40\\
& {\it NB}\,816 \tablenotemark{c}   & 4.49 & 0.59 & 25.68\\
\enddata
\tablenotetext{a}{The seeing denotes the FWHM of the point spread function.}
\tablenotetext{b}{Magnitude at which $50\%$ of detected sources have $\mathrm{S/N}\ge 5$.  Values are given for PSF-matched photometry within a $2.\!\arcsec0$ diameter aperture, after correcting for Galactic extinction.}
\tablenotetext{c}{These data were retrieved from the public data archive.}
\end{deluxetable}

The raw data were processed using the standard HSC pipeline version 6.7 (hscpipe6.7; \citealt{2018PASJ...70S...5B}), which is closely linked to the development of the software pipeline for the Large Synoptic Survey Telescope (LSST; \citealt{2010SPIE.7740E..15A,2015HiA....16..675J}).  The pipeline performs all the standard procedures including bias and dark subtraction, flat-fielding with dome flats, astrometric and photometric calibration, stacking, source detection, and multiband photometry.  The astrometric and photometric calibrations are based on the data of the Panoramic Survey Telescope and Rapid Response System (Pan-STARRS) 1 imaging survey \citep{2012ApJ...756..158S,2012ApJ...750...99T,2013ApJS..205...20M}.  Effective point spread functions (PSFs), which are listed in Table \ref{tb:observations}, are measured in the hscpipe procedure.  For each possible source detected through the procedure, the pipeline assigns pixel flags according to the status of pixels at the source position.  We used these flags to exclude those impacted by saturated pixels and other bad pixels.

We estimate the total magnitudes and colors of sources by using ``forced'' photometry with a $2.\!\arcsec0$ diameter aperture, in which multiband images are convolved with a PSF to match the seeing on the images to $0.\!\arcsec84$ in FWHM.  All of the magnitudes were corrected for Galactic extinction based on the dust map of \citet{1998ApJ...500..525S} with a 14\% recalibration \citep{2010ApJ...725.1175S} and a \citet{1999PASP..111...63F} reddening law.  We utilized the software dustmaps\footnote{https://github.com/gregreen/dustmaps} \citep{2018JOSS....3..695G} to get the reddening value for each source position.  Limiting magnitudes after extinction correction are summarized in Table \ref{tb:observations}.  

We limited the survey area to within $40\arcmin$ of the quasar position.  To exclude regions impacted by bright stars, we used bright-star masks that were built using {\it Gaia} DR2 \citep{2018A&A...616A..11G,2018A&A...616A...1G} to exclude sources down to $G_\mathrm{\it Gaia} < 18$ with magnitude-dependent apertures \citep{2018PASJ...70S...7C}.  We also masked extended sources brighter than $z_\mathrm{AB}\le19.5$ using elliptical apertures.  For masking, we utilized the software venice\footnote{https://github.com/jcoupon/venice}.  

To derive the effective survey area, we used a random-point catalog with a number density of $100~\mathrm{arcmin}^{-2}$ in the same geometry as the data.  We utilized the software maskUtils\footnote{https://github.com/jcoupon/maskUtils} that assigns the pixel flags for each random point according to the pixel status of the actual multiband images.  We can thus exclude random points with the equivalent criteria on the pixel flags as adopted in source selection.  After excluding those falling within the masked areas, the effective survey area is computed to be 1.255~deg$^2$ from the number count of the remaining random points.  The random-point catalog can also be used to derive the effective areas of arbitrary apertures placed within the survey field.

\section{Analysis and Results} \label{sec:analysis}

\subsection{Experimental design}
The goal of this experiment is to examine the presence of an overdensity or underdensity of galaxies in the vicinity of the extreme Ly$\alpha$ trough toward J0148$+$0600.  As tracers of the underlying density field, we will use LBGs which are selected to lie within the redshift interval of the trough ($5.5\le z\le5.9$).  We here provide an outline of our analysis. 
 
We first define a set of criteria on magnitudes and colors to construct a sample of LBG candidates by using model templates.  In doing so, we adjust the criteria such that the expected redshift distribution of the sample matches as well as possible the redshift range of the Ly$\alpha$ trough.  This procedure is described in Section \ref{sec:selection criteria}.  We then present the sample of LBG candidates and compare the number counts with the expectation in Section \ref{sec:LBG sample}

The sample of LBG candidates will undoubtedly contain some contaminants, i.e., objects that are not in fact an LBG at the required redshift.  These may either be objects with colors that mimic LBGs or objects with different colors that are scattered into the LBG color selection region by photometric errors.  In Section \ref{sec:contamination}, we discuss the origins of contamination and give a rough estimate of the level of stellar contamination.

The philosophy of this paper is not to try to accurately determine the number of foreground contaminants because our primary goal is to measure the relative density around the quasar position, assuming that the distribution of contaminants is independent of the presence of the background quasar.  The outer region of the HSC field is used to compare  the densities self-consistently.  Of course, possible spatial effects in the contamination rate due to instrumental characteristics will be more serious, not least the noise level (flux errors) increasing with radius from the center.  The spatial variations in sensitivity affect not only the detection in the $z$-band but also whether a source meets the ($i2-z$) color criteria and the nondetection in $r2$.  Consequently, spatial variations arise in the selection function and induce biases in the surface density measurements of LBG candidates in a spatial-dependent way.  These must be certainly evaluated.

To evaluate these effects, in Section \ref{sec:mock}, we use a deeper HSC dataset to simulate our observations and analysis while adopting an intrinsically spatially uniform distribution of objects.  Using a number of Monte Carlo realizations, we first obtain rough estimates of the expected numbers of LBG candidates and possible contaminants and compare these with the data.  We then evaluate the surface density variations in LBG candidates that arise purely due to the spatial variations in noise.  The goal of our experiment is to search for real density variations in the field that cannot be ascribed to spatial variations in the contamination rate due to spatial variations in the instrumental sensitivity.

Lastly, we can in principle make a test for the rare-source model \citep{2015MNRAS.453.2943C,2017MNRAS.465.3429C} by searching for luminous quasars in the large cylindrical survey volume along the Ly$\alpha$ trough.  This test is independent of the LBG surface density measurements.  The test must be limited to a search for unambiguous cases because it is usually challenging to distinguish quasars from Galactic stars for their point-source-like light profile and similar colors.

\subsection{LBG Selection criteria} \label{sec:selection criteria}

\begin{figure}[ttbp] 
   \centering
   \includegraphics[width=3.5in]{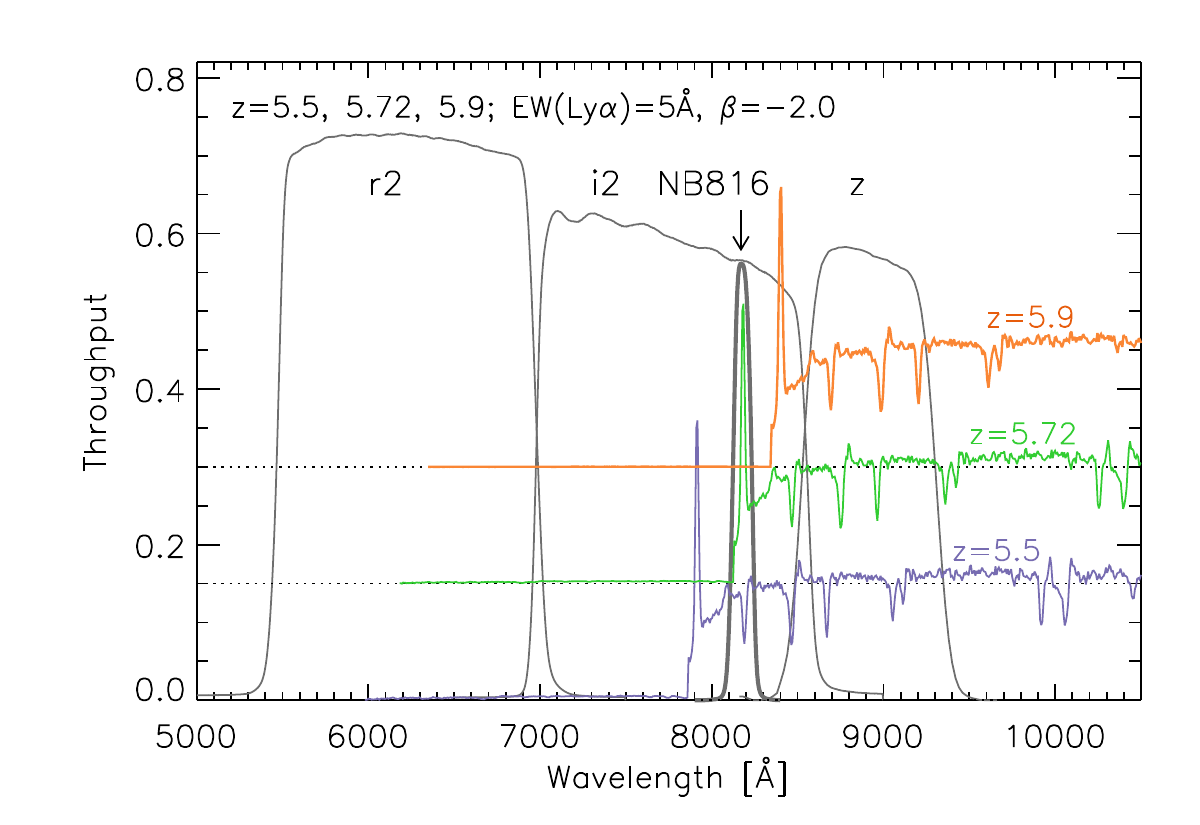} 
   \caption{Throughput of the HSC filters, $r2$, $i2$, $z$ (solid lines), and {\it NB}\,816 (thick solid line).  LBG template spectra of $\mathrm{EW_0(Ly\alpha)}=5~\mathrm{\AA}$ and $\beta=-2.0$ at $z=5.5$, 5.72, and 5.9 are shown by colored lines.  At these redshifts, the ($i2-z$) color is sensitive to the fraction of flux redder than the break falling inside the $i2$ bandpass and thus can be used as an indicator of redshift.  The narrowband {\it NB}\,816 filter used in \citet{2018ApJ...863...92B} is sensitive to LAEs within a narrow redshift range ($\Delta z\approx 0.1$) centered at $z\approx5.7$.}
   \label{fig:LBGselection}
\end{figure}

In this work, we will construct a sample of high-$z$ LBGs using the $r2$, $i2$, and $z$ filters by selecting sources whose UV continuum is sampled in the $z$-band and the break due to Ly$\alpha$ forest absorption falls in the $i2$ band, as shown in Figure \ref{fig:LBGselection}.  The ($i2-z$) color is used to constrain the redshift, and the $r2$ data are used to remove low-redshift interlopers.  Our strategy is similar to the method of \citet{2014MNRAS.442..946D}, but was modified for the use of the HSC filters and to select LBGs from the redshift ranges of interest.  

We define the selection criteria by using template spectra of high-$z$ LBGs which are generated based on a composite spectrum of LBGs at $z\approx2.94$ from \citet{2003ApJ...588...65S}.  The spectrum is modified to simulate various UV spectral slopes $\beta$ ($f_\lambda \propto \lambda^\beta$), Ly$\alpha$ equivalent widths (EWs), and effects of the Ly$\alpha$/Ly$\beta$ forest absorption following the measurements of Ly$\alpha$ optical depths as a function of redshift \citep{2006AJ....132..117F,2015MNRAS.447.3402B,2018MNRAS.479.1055B}.

\begin{figure}[tbp] 
   \includegraphics[width=3.5in]{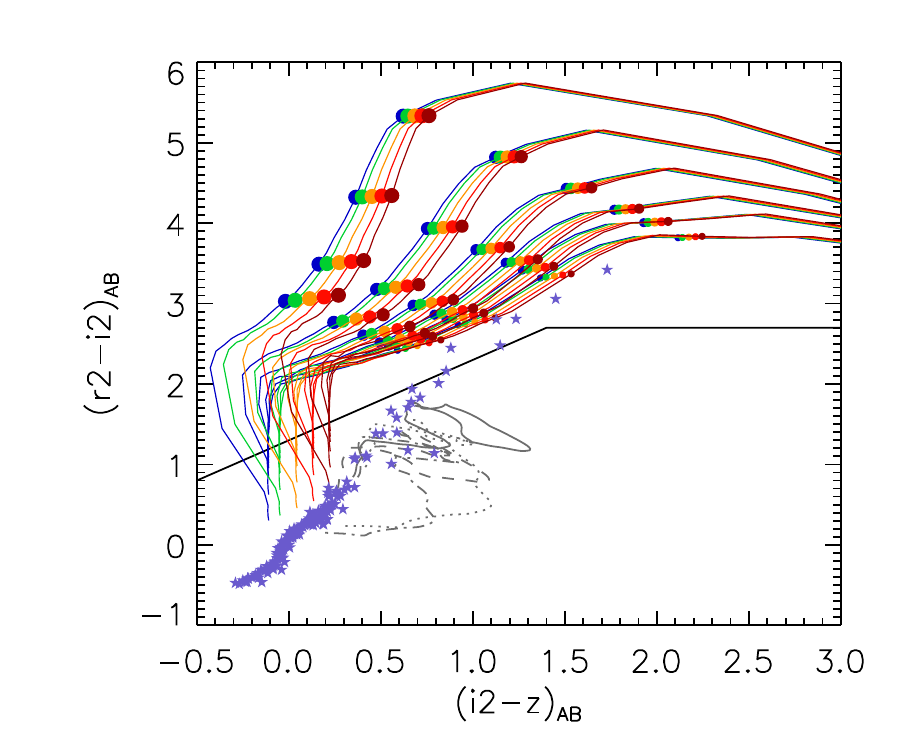}
   \includegraphics[width=3.5in]{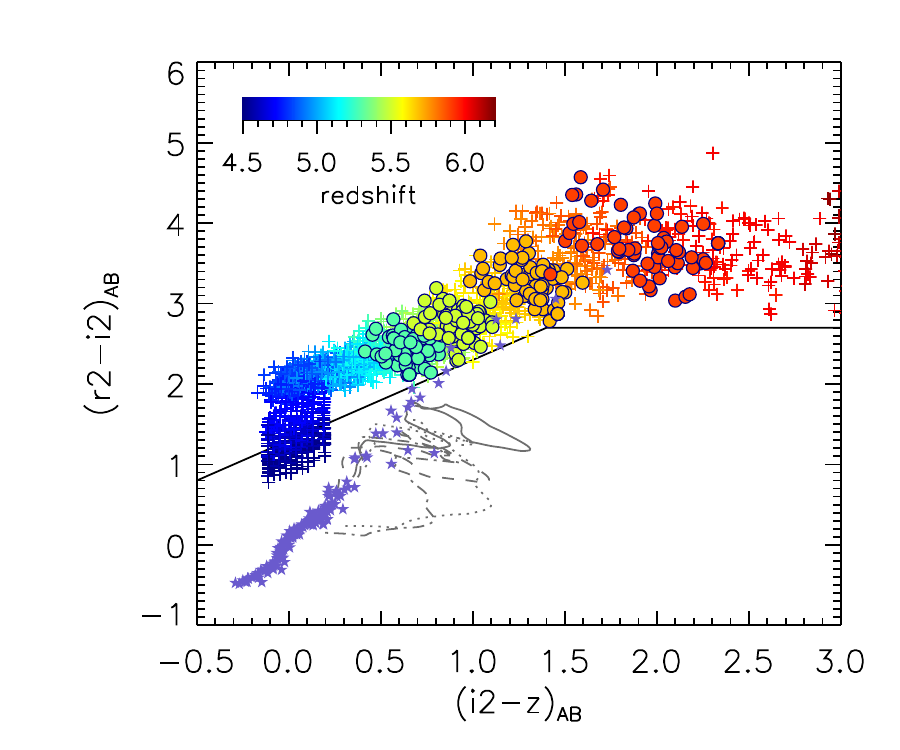}
      \caption{Upper panel: color-color diagram ($i2-z$) versus ($r2-i2$).  Colored solid lines indicate color tracks for LBG templates from redshift $z=4.5$ to 6.2 from the lower left to the upper right.  Different groups of tracks correspond to different $\mathrm{EW_0}$(Ly$\alpha$) of 0, 10, 20, 40, 80, and 160~{\AA} from the bottom to top.  The five colors correspond to different UV slopes $\beta$ of $=-3$ to $-0.8$ from blue to red.  The colors at $z=5.3$, 5.5, 5.7, and $5.9$ are highlighted by circles.  The gray lines indicate the color tracks for elliptical (5 Gyr solid; 2 Gyr dashed), S0 (dotted), and Sb (dotted-dashed) galaxies redshifted from $z=0.5$ to $2.5$.  Purple stars indicate the colors of Galactic O--M stars based on low-resolution spectra from \citet{1983ApJS...52..121G}.  The black solid line marks the threshold in ($r2-i2$) for high-$z$ LBGs.  Lower panel: same as the upper panel, but showing the template colors with the effects of fluctuating Ly$\alpha$ forest absorption.  EW$_0$(Ly$\alpha$) is limited to $\le 30~\mathrm{\AA}$.  Symbols are color-coded by input redshift and are highlighted with circles at $z=5.3$, $5.5$, $5.7$, and 5.9.}
   \label{fig:iz_vs_ri_templates}
\end{figure}

In the upper panel of Figure \ref{fig:iz_vs_ri_templates}, we show the redshift evolution of the ($i2-z$) and ($r2-i2$) colors from $z=4.5$ to 6.2 with different rest-frame $\mathrm{EW_0}(\mathrm{Ly\alpha})$ and continuum slope $\beta$.  The colors at $z=5.3, 5.5, 5.7$, and $5.9$ are highlighted by circles.  These tracks indicate that ($i2-z$) is sensitive to redshift.  The variations in $\beta$ have a minimal impact on these colors.  In contrast, the Ly$\alpha$ emission could significantly affect the colors of the galaxies.  Strong Ly$\alpha$ emission make the ($i2-z$) color bluer, and the difference in ($i2-r2$) reaches $\sim 1~\mathrm{mag}$ between $\mathrm{EW_0(Ly\alpha)}=0~\mathrm{\AA}$ and $80~\mathrm{\AA}$.  High-$z$ LAE surveys suggest that, though most LBGs at $z\sim6$ display Ly$\alpha$ emission, strong ($\mathrm{EW_0(Ly\alpha)}\gtrsim40~\mathrm{\AA}$) LAEs do not dominate amongst moderately bright LBGs \citep{2011ApJ...734..119K,2012ApJ...744...83O}.

In the lower panel of Figure \ref{fig:iz_vs_ri_templates}, we show the template colors that include the effects of fluctuating Ly$\alpha$ forest absorption.  The probability distribution of the Ly$\alpha$ optical depth follows fits provided by \citet{2018MNRAS.479.1055B}.  Here, EW$_0$(Ly$\alpha$) is limited to $\le30~\mathrm{\AA}$ as we consider relatively massive LBGs with moderate Ly$\alpha$ emission.  Symbols are color-coded by input redshifts in the range $4.5\le z\le 6.2$.  The highlighted circles at $z=5.3,~5.5,~5.7$, and 5.9 show that the intrinsic ($i2-z$) colors are well separated with a redshift difference of $\Delta z \sim 0.2$, i.e., this corresponds to the redshift uncertainty without photometric errors.  

Based on these template colors and tests described below, we defined the color criteria to select LBGs from the redshift range of interest ($5.5\lesssim z\lesssim5.9$) as follows:
\begin{eqnarray}
&&1.0\le (i2-z)_\mathrm{AB} \le 2.3,\text{ and} \nonumber \\
&&\left[\left( \mathrm{S/N}(r2) < 2 \textrm{ and } r2_\mathrm{AB} \ge 27.9\right)\text{ or }\right.  \\
&&\left. (r2-i2)_\mathrm{AB} \ge \text{ min}\!\left\{2.7,~(i2-z)_\mathrm{AB}+1.3\right\} \right]. \nonumber
\end{eqnarray}
We also limit sources to having both $24.0\le z_\mathrm{AB}\le 25.3$ and signal-to-noise ratio $(z_\mathrm{AB})\ge5.0$.  The faint-end limit corresponds to an absolute UV magnitude $M_\mathrm{UV}\approx-21.4$ at $z=5.7$.  The bright-end limit is imposed to reduce the Galactic stellar contamination.  For the same purpose, we further excluded point-like sources with $z_\mathrm{AB} < 24.5$, which show a radial light profile that is not extended more than the PSF\footnote{Point-like sources were identified with {\tt base\_ClassificationExtendedness\_value~$=0$.}} ($\approx9\%$).  Finally, each source satisfying these criteria was visually inspected to reject hot pixels, unflipped cosmic rays, moving objects, and noise features in the outskirts of bright objects.

\begin{figure}[tbp] 
   \centering   
   \includegraphics[width=3.5in]{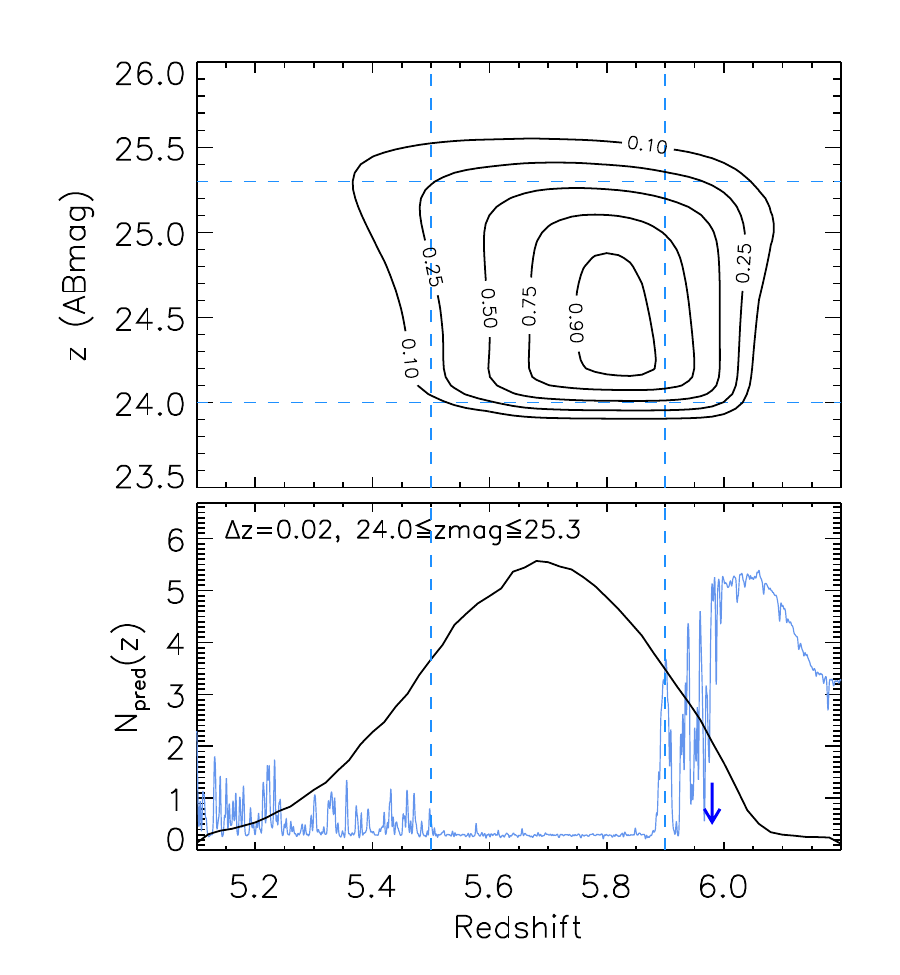}
   \caption{Upper panel: expected completeness $C(z, m_z)$ as a function of redshift and $z_\mathrm{AB}$ magnitude ($m_z$).  The contours indicate probabilities that a simulated LBG at a given redshift and true $z$ magnitude is detected and satisfies our selection criteria.  The horizontal and vertical dotted lines indicate the limiting magnitudes ($24.0\le z_\mathrm{AB}\le25.3$) and the redshift range of interest ($5.5\le z \le 5.9$).  Lower panel: expected redshift distribution computed from the completeness and the UV luminosity functions.  The blue line indicates the spectrum of the quasar J0148$+$0600 in arbitrary units whose observed wavelength is converted into redshift based on the wavelength of the Ly$\alpha$ emission line.  The quasar redshift $z=5.98$ is marked by an arrow.  The expected redshift selection function has a broad peak at the center of the Ly$\alpha$ trough and spreads to cover the entire redshift range of the trough across $5.5\le z \le 5.9$.}
   \label{fig:completeness}
\end{figure}

The above criteria were tested and iteratively optimized as follows.  Taking a set of galaxy templates with various $\beta$ and $\mathrm{EW_0}(\mathrm{Ly}\alpha)$ ($\le 30~\mathrm{\AA}$) across a wide redshift range, we rescale their total magnitudes and distort their true $r2$, $i2$, and $z$ magnitudes by adding Gaussian noises to mimic the quality of the HSC data.  We then compute the selection probability, or completeness $C(z, m_z)$, as a function of redshift and $z$ magnitude ($m_z$) through Monte Carlo realizations.  Figure \ref{fig:completeness} shows in the upper panel the completeness $C(z, m_z)$, which denotes the chance that a simulated high-$z$ LBG at given redshift and $z$ magnitude meets the above selection criteria after observational noise is added.  Note that the $y$-axis denotes the true $z$ magnitude and thus the outskirt of the contours is extended beyond the limits on $z$ magnitude because of the effects of noise.  

To predict the redshift distribution of those that meet the criteria, we integrate $C(z, m_z)$ along the magnitude axis while adopting the UV luminosity function interpolated at each redshift \citep{2015ApJ...803...34B}.  The result is shown in the lower panel of Figure \ref{fig:completeness}, in comparison with the Ly$\alpha$ forest spectrum of J0148$+$0600 that is converted to redshift space based on the Ly$\alpha$ wavelength.  The expected redshift distribution has a broad peak at the center of the trough and covers the full length across $5.5\le z\le 5.9$.  We find that the expected contribution from the outskirts ($z<5.5$ and $z>5.9$) is 33\%.  This prediction is made for the standard luminosity function, and the actual redshift distribution of real LBGs in the J0148$+$0600 field would be affected by a possible over- or underdensity associated with the trough if it really exists.

A concern is that the spatial distribution of LBG candidates may be affected by the local environment of the background quasar itself as the high-$z$ wing of the expected redshift distribution extends beyond the quasar redshift (Figure \ref{fig:completeness}) and thus we could bring forward into the sample, $z\gtrsim6$ LBGs, i.e., those around the quasar itself.  A possible overdensity around the luminous quasar \citep[e.g.,][]{2014A&A...568A...1M} might enhance the galaxy count.  However, this possible effect will not impact our conclusions because we will in fact find an underdensity, which is in the opposite sense to this effect.  On the other hand, a possible proximity effect should also be negligible because, if it really exists, it could affect only those in a much fainter regime ($M_\mathrm{UV}\gtrsim17$; \citealt{2019ApJ...870...45U}) than our LBG candidates.

One might also worry that the high opacity in the vicinity of the trough could systematically change the colors of the galaxies and induce a bias in the resultant redshift distribution of LBGs extracted with given color criteria.  Concretely, a higher Ly$\alpha$ forest absorption makes the ($i2-z$) color redder and thus the average redshift of those selected within a given ($i2-z$) range may be biased toward a lower value.  We have checked that the effects are negligible by simulating highly opaque Ly$\alpha$ forest absorption in the templates.

\subsection{LBG candidate sample} \label{sec:LBG sample}

\begin{figure}[tbp] 
   \includegraphics[width=3.5in]{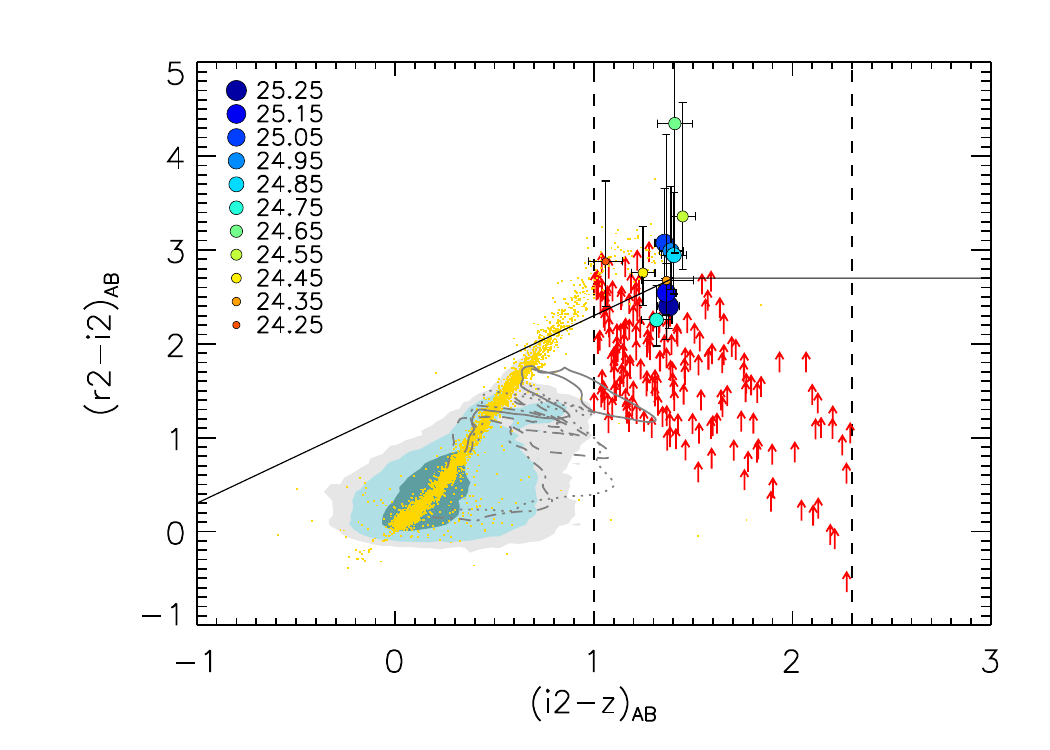}
      \caption{The ($i2-z$) versus ($r2-i2$) diagram for all sources in the catalog.  The red upward-pointing arrows indicate nominal $(i2-z)$ colors and lower limits on $(r2-i2)$ for the LBG candidates.  The circles indicate the colors computed from the total fluxes of subsamples binned by $z$ magnitude with different sizes and colors, according to the legend on the left.  The contours correspond to the full catalog of detections at $z_\mathrm{AB}\le25.3$ and contain 50\%, 90\%, and 97\% of the sources.  The gray lines indicate the color tracks of $z\sim1\textrm{--}2$ galaxies (same as in Figure \ref{fig:iz_vs_ri_templates}).  Yellow dots indicate the candidates of bright stars ($20\le z_\mathrm{AB}<22$) in the catalog.  Vertical dashed lines indicate the selection limits on $(i2-z)$.}
   \label{fig:iz_vs_ri}
\end{figure}

According to the selection criteria above, we selected 185 sources as potential LBGs at $z\sim5.5\textrm{--}5.9$, along with 25 relatively bright ($24\le z_\mathrm{AB}< 24.5$) point-like objects satisfying the color criteria identified as possible stars.  These are hereafter excluded.  The mean surface density is $\bar{\Sigma}_\mathrm{LBGc} = 0.041~\mathrm{arcmin}^{-2}$ ($\bar{\Sigma}$ denotes the mean surface density across the field and the subscript ``LBGc'' stands for ``LBG candidates'').  Figure \ref{fig:iz_vs_ri} shows the $(i2-z)$ versus $(r2-i2)$ color diagram.  LBG candidates are highlighted by upward-pointing arrows, which indicate the $1\sigma$ lower limits of the $(r2-i2)$ color.  For reference, the contours contain 50\%, 90\%, and 97\% of the sources of the full catalog of detections at $z_\mathrm{AB}\le 25.3$.  Yellow dots indicate point-like sources likely to be Galactic stars at $20\le z_\mathrm{psf} \le 23$, where $z_\mathrm{psf}$ denotes the forced PSF magnitude.  The LBG candidates are reasonably separated from the bulk of the full sample and stars.  The $(i2-z)$ colors of the LBG candidates are spread across the selection interval of $1.0\le (i2-z)_\mathrm{AB} \le 2.3$ while being skewed toward the lower boundary.

It is clear that most of the LBG candidates have 1$\sigma$ lower limits of their $(r2-i2)$ colors that are substantially below the nominal color selection, as indicated by the solid line in Figure \ref{fig:iz_vs_ri}.  We first check, by stacking, whether these objects have average fluxes that are consistent with the $(r2-i2)$ colors of the LBG templates.  We computed total colors by summing the nominal flux measurements of the LBG candidates in bins of $z$ magnitude with a bin size of $\Delta z_\mathrm{AB}=0.1$.  In Figure \ref{fig:iz_vs_ri}, these stacked colors are indicated by circles color-coded according to the $z$ magnitude.  These circles are all located above or on the boundary that corresponds to the lower $(r2-i2)$ limit for high-$z$ LBGs.  This is reassuringly consistent with the idea that most of the LBG candidates may truly lie in the color region of LBGs and that a relatively small number of low-$z$ contaminants are included.

\begin{figure}[tbp]
   \centering
      \includegraphics[width=3.5in]{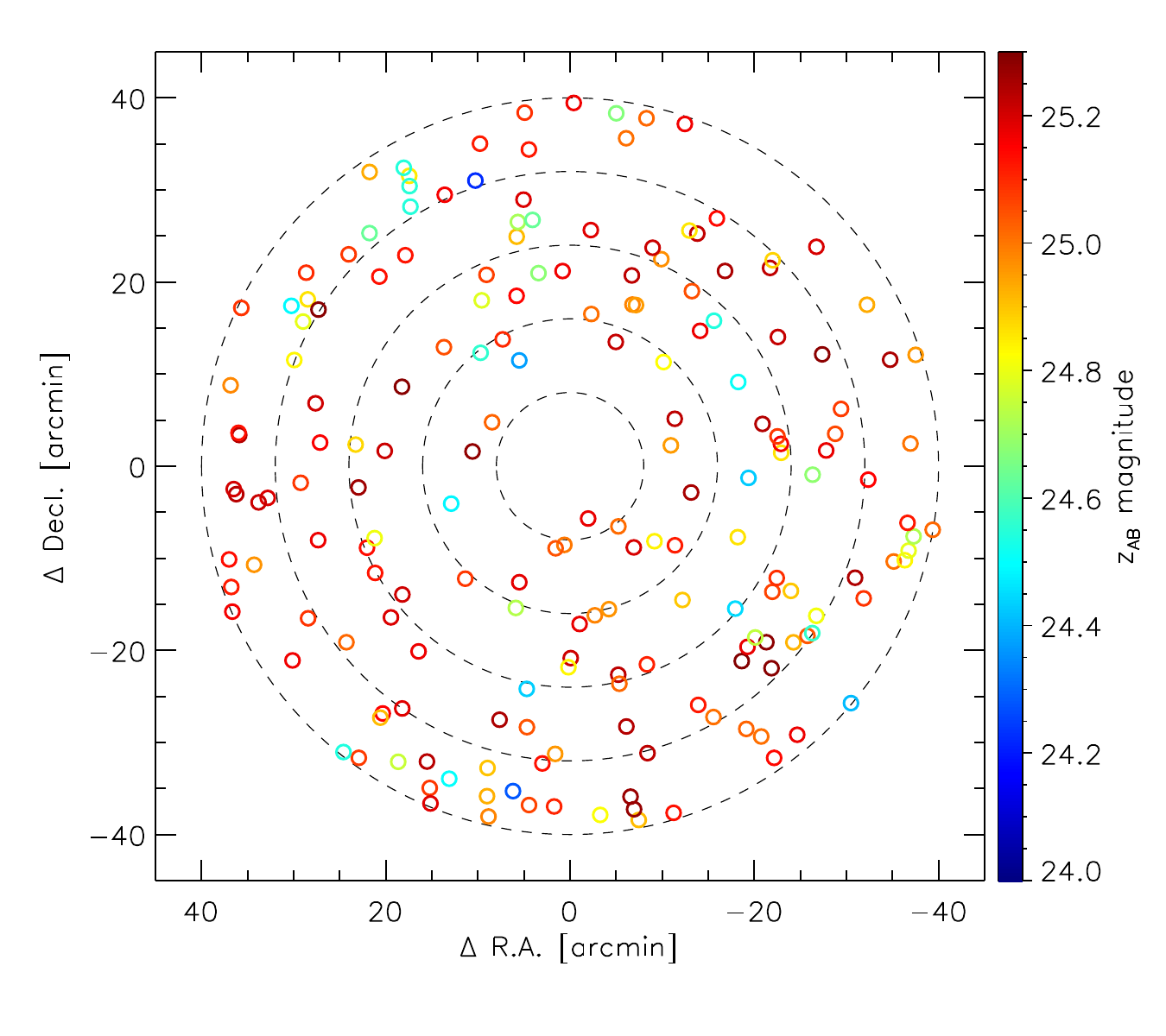}
   \caption{Distribution of LBG candidates on the sky.  The field is centered on the position of the quasar J0148+0600.  Symbols are color-coded by the $z$ magnitude according to the color scale on the right.  The dashed lines show concentric circles of radii in increments of $8\arcmin$  from the field center.}
   \label{fig:LBG_maps}
\end{figure}

As noted above, our primary interest is in the spatial distribution of objects on the sky.  The spatial distribution of the 185 LBG candidates is shown in Figure \ref{fig:LBG_maps}.  Symbols are color-coded according to their $z$ magnitude.  A possible deficit of LBGs is immediately seen in the vicinity of the position of J0148$+$0600, i.e., the field center.  Before drawing any conclusions, however, we must address the question of whether spatial variations in the contamination rate could produce similar effects.

\begin{figure}[tbp]
   \centering
   \includegraphics[width=3.5in]{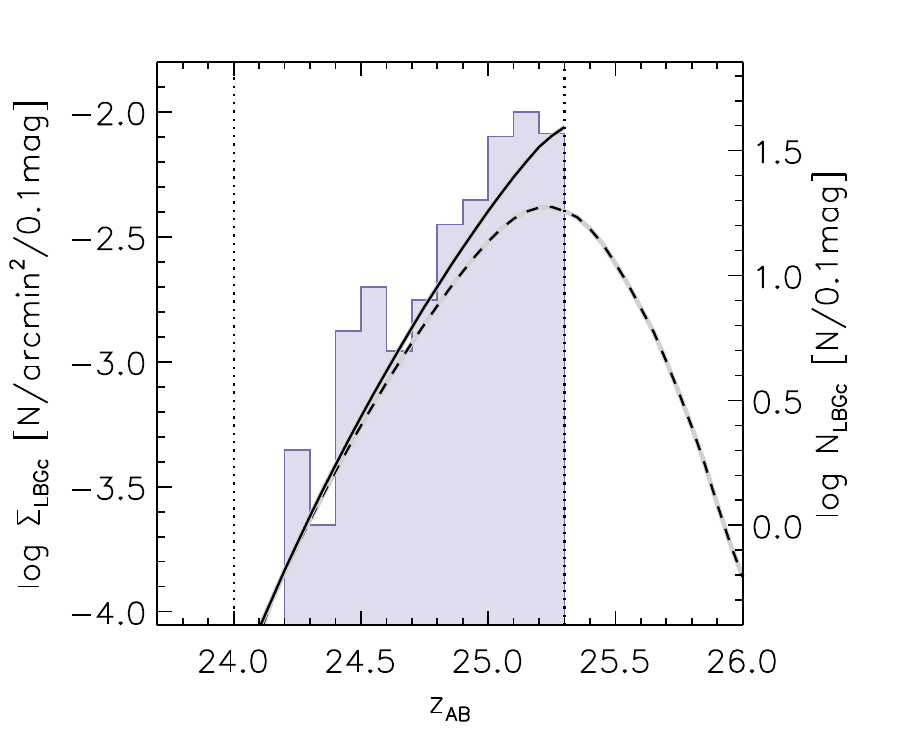}
   \caption{Surface density of LBG candidates as a function of $z$ magnitude is shown by a filled histogram.  The $y$-axis on the right-hand side indicates the corresponding number count in log scale.  The vertical dotted line indicates the lower and upper limits for selection.  The solid and dashed curve indicates the predicted distribution of the observed and true $z$ magnitudes, respectively, which are computed based on UV luminosity functions. The right wing of the true $z_\mathrm{AB}$ distribution ($z_\mathrm{AB}>25.3$) corresponds to candidates that are intrinsically fainter than the limit but scattered into the selection region due to photometric errors.}
   \label{fig:nofm}
\end{figure}

Before doing so, in Figure \ref{fig:nofm}, we plot the surface density of LBG candidates as a function of $z$ magnitude. For comparison, we derived the expected surface density against the $z$ magnitude, accounting for the observational effects by integrating the completeness $C(z, m_z)$ (Figure \ref{fig:completeness}) along the redshift axis while adopting the UV luminosity function interpolated at a given redshift \citep{2015ApJ...803...34B}.  First, the shape of the distribution (histogram) is in good agreement with the predicted distribution of the observed $z$ magnitude  (solid line).  However, the observed mean surface density ($0.041~\mathrm{arcmin}^{-2}$) is found to be higher than the predicted value of $0.032~\mathrm{arcmin}^{-2}$ with the same $z_\mathrm{AB}$ limits.  This suggests that, if the survey field is representative of the cosmic average, the sample contains $\approx 23\%$ contamination (or $\sim 40$ among 185 candidates).  In the following subsection, we will indeed find that our LBG candidate sample is expected to contain significant contamination, and the fact that the observed mean density of candidates is higher than the prediction is not to be regarded as a sign of an overdensity of LBGs in this field.

\subsection{Contamination} \label{sec:contamination}

The contamination in the color selection for high-$z$ LBGs comes mainly from Galactic M stars, brown dwarfs, and red galaxies at intermediate redshifts.  We first assess the expected number of stellar contaminants.  In Figure \ref{fig:iz_vs_ri_templates}, we show Galactic O--M stars whose colors are computed based on spectra given by \citet{1983ApJS...52..121G} adopting the HSC filter functions.  The sequence reaches the same color region occupied by high-$z$ LBGs.  In addition, L and T brown dwarfs could be a source of contamination because they have an ($i2-z$) color as red as LBGs and are usually undetectable in $r2$.  By using a compilation of public spectra of low-mass stars\footnote{We retrieved stellar spectra from \citet{1983ApJS...52..121G}, the L and T dwarf data archive built by Sandy Leggett (\url{http://staff.gemini.edu/~sleggett/LTdata.html}), and the SpeX Prism Spectral Libraries built by Adam Burgasser (\url{http://pono.ucsd.edu/~adam/browndwarfs/spexprism/}).}, we find that dwarfs in the spectral type range of M7--T5 are in practical terms indistinguishable from LBGs in the ($r2$, $i2$, $z$) color space.

We estimated the number density of these types of dwarfs using the study of \citet{2008A&A...488..181C}, where the absolute magnitude and the local number density measurement are available for each of the spectral types of dwarfs.  With a simple Galaxy model in the literature, we computed the expected number counts of M7--T5 dwarfs in the direction of J0148$+$0600.  In doing so, for each spectral type, we computed the expected total number within a pencil light cone across the distance range at which they are to be observed with an apparent $z$ magnitude between $24.5\le z_\mathrm{AB} \le25.3$.  Summing up the spectral types of M7--T5, we found the expected number of stellar contaminants to be $\sim60$ ($0.013~\mathrm{arcmin}^{-2}$) in the survey area ($1.25~\mathrm{deg}^2$) with a major contribution from late-M stars.  

The second contamination source may be evolved galaxies at intermediate redshifts ($z\sim1\textrm{--}2$) whose significant $4000~\mathrm{\AA}$ break could mimic the Ly$\alpha$ break in high-$z$ LBGs.  In Figure \ref{fig:iz_vs_ri_templates}, we show color tracks of elliptical, S0, and Sb galaxies redshifted from $z=0.5$ to 2.5 based on template spectra of \citet{2007ApJ...663...81P}\footnote{The SWIRE template library: \url{http://www.iasf-milano.inaf.it/~polletta/templates/swire_templates.html}}.  The reddest colors of these low-$z$ interlopers still differ from the region occupied by LBGs, but they may contaminate the LBG sample because of photometric errors and the fact that the $r2$ depth is not sufficient to exclude these objects (see Figure \ref{fig:iz_vs_ri}).  A specific concern is that these galaxies may be spatially correlated on the sky, whereas the stellar contaminants can be assumed to be randomly distributed across the field.

We repeat that the simple test that uses the stacked colors of LBG candidates suggests a relatively small degree of contamination from low-$z$ red galaxies (Section \ref{sec:LBG sample}).  In addition, the estimated level of stellar contamination is in agreement with the overdetection relative to the prediction from the UV luminosity function (see Section \ref{sec:LBG sample}).  This agreement also suggests that other types of contamination likely have a minor contribution.  Unfortunately, however, we have no way to estimate the number of such faint ($z_\mathrm{AB}\ge24$) red galaxies at $z\sim1\textrm{--}2$ theoretically.  However, there is no reason to suppose that the spatial distribution of the foreground galaxy contaminants would be related to the position of the background quasar.  

Another probably more serious concern is that the detection rates of real LBGs and contaminants may be spatially biased, likely in different ways, due to spatially varying photometric errors.  We will therefore make further tests in the following section to estimate the degree of contamination and the spatial distortions of the selection function across the whole field. 

\subsection{Tests with the COSMOS ultradeep data} \label{sec:mock}

\begin{figure}[tbp] 
   \centering
   \includegraphics[width=3.5in]{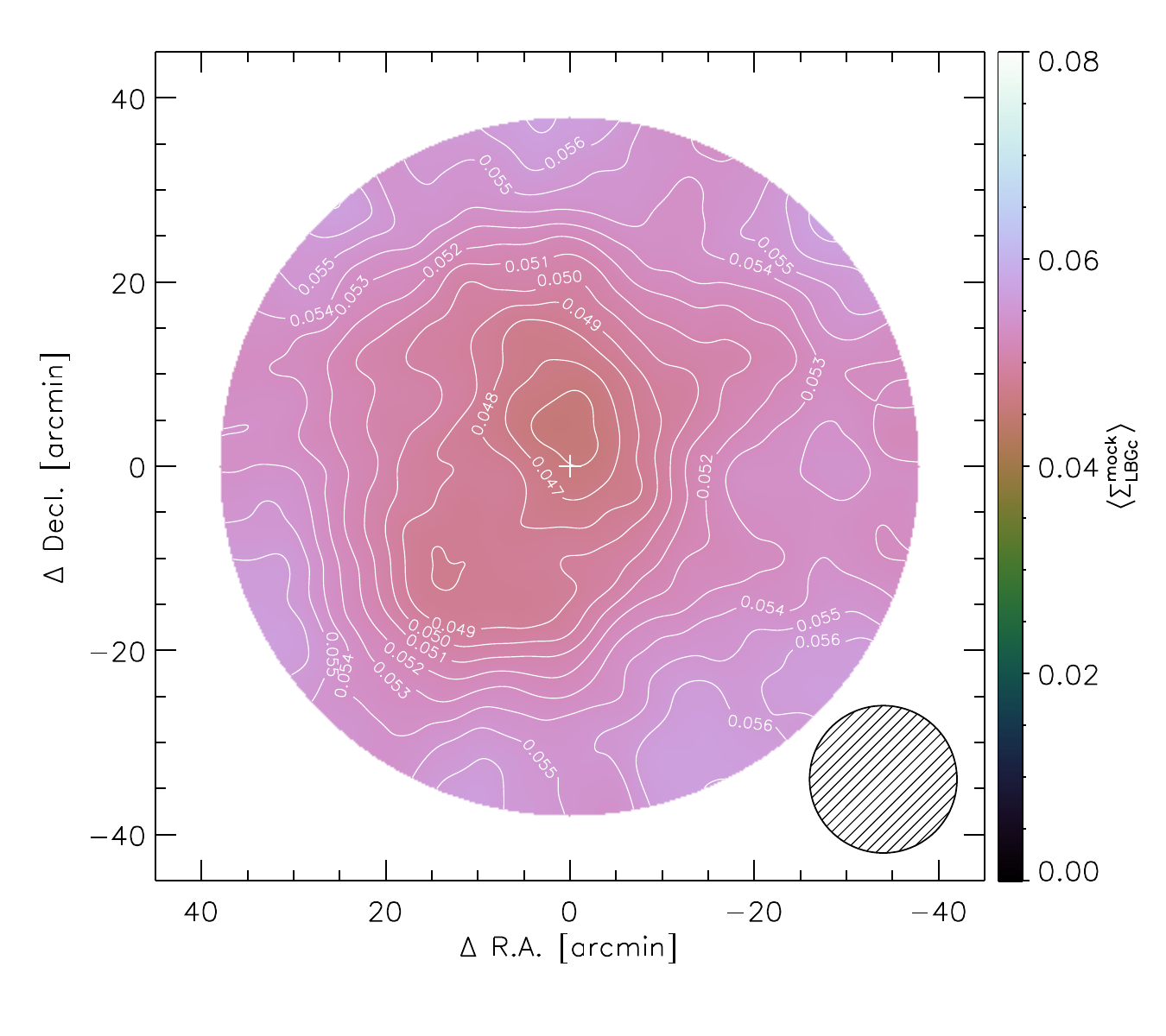}
   \caption{The expected mean surface density $\left< \Sigma^\mathrm{mock}_\mathrm{LBGc} (\alpha, \delta) \right>$ is shown with the same color scale as Figure \ref{fig:LBG_densitymap}.  Contours are overplotted for visibility.  The variations are $\lesssim 15\%$ against the mean value across the field.}
   \label{fig:map_predicted}
\end{figure}

In this section, we estimate possible spatial effects due to spatially varying photometric noise.  For these purposes, we used mock catalogs that were created with a subset of real HSC data in the ultradeep (UD) COSMOS field from the second public data release of the HSC-Subaru Strategic Program (SSP; \citealt{2019PASJ...71..114A}).  The UD-COSMOS data contains all five band ($grizy$) photometry and is sufficiently deep that an accurate classification among high-$z$ LBGs, Galactic stars, and foreground galaxies can be made.  Before creating mocks, we categorized all the UD-COSMOS objects into either high-z LBGs, low-$z$ galaxies, or Galactic stars.  In particular, we utilized the $y$-band photometry, which is not available in our HSC data, to more accurately distinguish brown dwarfs because they show redder $(z-y)_\mathrm{AB}$ $(\gtrsim0.6$) than LBGs.

We created mock catalogs, each of which corresponds to a single realization of our HSC observation but imposes an intrinsically spatially uniform distribution of all types of sources, as follows.  For each UD-COSMOS object, we assigned a random sky position ($\alpha, \delta$) and distorted their photometry by using the flux errors of a real detection in our catalog that is nearest to the random point.  In doing so, the number density of the UD-COSMOS catalog was preserved, i.e., the total number of sources was rescaled to match the survey area of our observation.  This procedure allows us to simulate our observation with a uniform source distribution while reflecting the spatial variations in noise.  We then adopted the same selection criteria for each mock catalog.  Through the selection, we excluded point-like objects at $z_\mathrm{AB}<24.5$, which are supposed to be stars, as done in our data.  Finally, we averaged the results over an ensemble of 500 mock catalogs.

First, we attempt to test the consistency between our data and the mocks.  The number of mock LBG candidates is found to be $N^\mathrm{mock}_\mathrm{LBGc}=238\pm13$ (``LBGc'' stands for ``LBG candidates''), which is in reasonable agreement with the observed number (185).  This already suggests strongly that our LBG candidate sample is sensible and that the UD-COSMOS mock catalogs are realistic.  We look into the details.  According to our classification of the UD-COSMOS sources, we found the fraction of non-LBG sources (i.e., the contamination from stars and red galaxies) to be $\approx48\%$ (or $N^\mathrm{mock}_\mathrm{contam}=115\pm9$).  Of these, stars and dwarfs account for $\approx46\%$ (or $N^\mathrm{mock}_\mathrm{stars}=53\pm6$).  The number of stellar contaminants is in good agreement with the expectation from the Galaxy model made in Section \ref{sec:contamination}.  The other $\approx60$ contaminants are expected to be red galaxies at $z\sim1\textrm{--}2$.  The relatively small fraction of red galaxy contamination ($\sim26\%$ of the total set of candidates) is good news, because it reduces the possible effects due to the galaxy autocorrelation as mentioned in Section \ref{sec:contamination}.  Lastly, the number of bright possible stars that were excluded before the color selection is found to be $39\pm4$, which is roughly $2\times$ higher than that of our data (25).  

After all, we are able to argue that these values from the data, the mocks, and the theoretical predictions (see Section \ref{sec:contamination}) are all reasonably consistent.  It is therefore sensible to expect a similar level of contamination ($\sim50\%$) for our LBG candidate sample.  Our philosophy is, however, not to try to determine very accurately the absolute abundance of LBGs, but rather to compare self-consistently the relative density around the quasar position to that of the outer region.  Therefore, the accuracy of the contamination fraction determination is not a critical issue, and our goal is achievable at some level even though the contamination fraction is high, because it can be assumed that the stellar and foreground galaxy contaminants are independent of the background quasar position.  

However, possible spatial biases due to the spatial variations in the noise could affect the conclusions.  To address the concern, we created a map of the ensemble mean of the LBG candidate surface densities $\left< \Sigma^\mathrm{mock}_\mathrm{LBGc} (\alpha, \delta) \right>$ (the brackets denote the ensemble average of the mock catalogs).  This is shown in Figure \ref{fig:map_predicted} with the smoothing scale and the color scale that are matched to those of the result presented in the following section.   In the figure, the presence of some spatial variations within $\lesssim15\%$ is clearly seen with an overall trend of increasing mean surface density with radius.  Again, note that this variation arises purely due to the spatially varying noise, because an intrinsically uniform distribution was adopted in the mocks.  The typical radial trend in $\left< \Sigma^\mathrm{mock}_\mathrm{LBGc} (\alpha, \delta) \right>$ arises due to the radial-dependent detection rate of contaminants.  In contrast, the spatial variation in the surface density of true LBGs is found to be almost uniform in the mocks.

The spatial biases in the contaminants could induce a spurious underdensity of LBG candidates near the field center and thus the relative fluctuations of the observed surface density must be evaluated accounting for these spatial effects.  In the following subsection, we evaluate whether the observed deficit in LBG candidates around the quasar position is meaningful in comparison with the $\left< \Sigma^\mathrm{mock}_\mathrm{LBGc} (\alpha, \delta) \right>$ map.

\subsection{Spatial distribution of LBG candidates}

\begin{figure}[tbp] 
   \centering
      \includegraphics[width=3.5in]{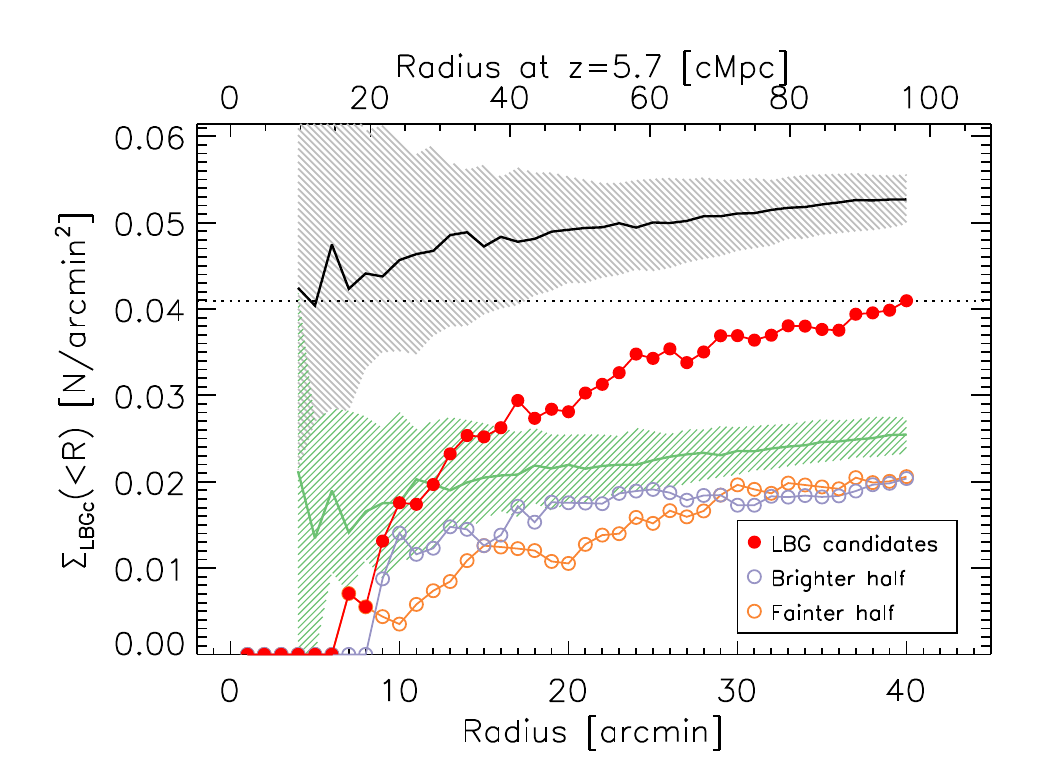}
      \includegraphics[width=3.5in]{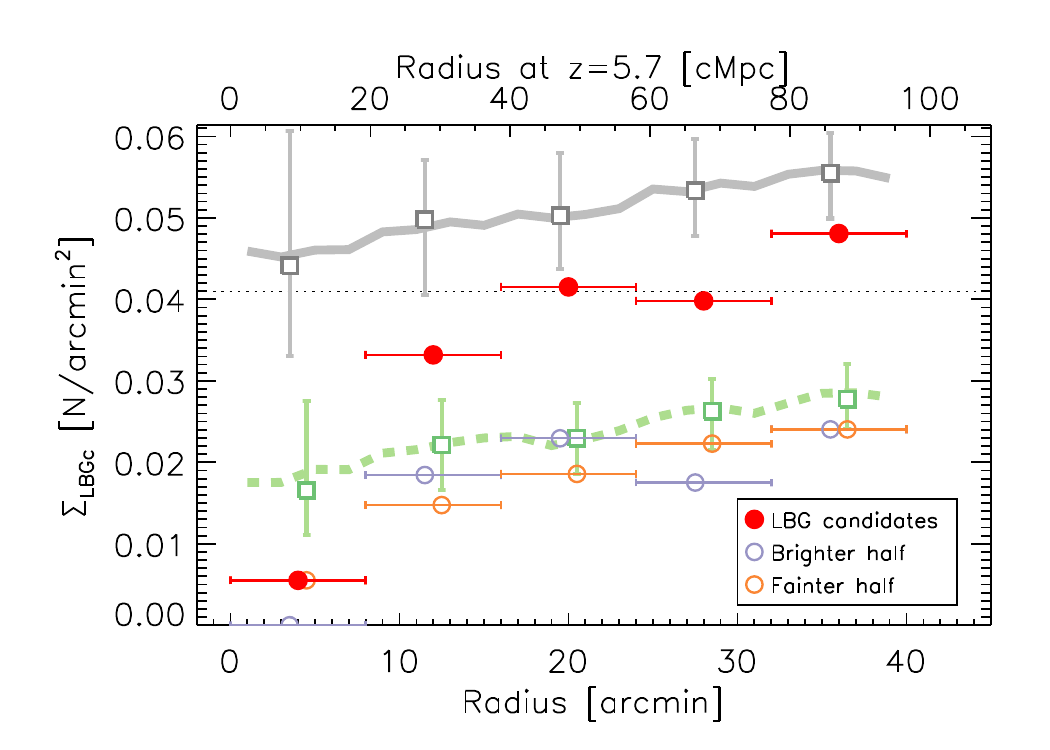}
   \caption{
   Upper panel: surface density of LBG candidates within a given radius from the quasar position in steps of $1\arcmin$.  The red filled circles indicate the results in the J0148$+$0600 field.  The open purple and orange circles correspond, respectively, to the brighter and fainter half of the LBG candidate sample.  The horizontal dotted line indicates the mean surface density ($0.041~\mathrm{arcmin}^{-2}$) across the field.  The solid lines indicate the ensemble average of the mocks for all detections that meet the LBG selection criteria (black) and possible contaminants (green).  Corresponding hatched regions are the 16th--84th percentiles of 500 realizations of the mock.  Lower panel: surface densities averaged over annular bins of $8\arcmin$ (red circles).  The horizontal error bars correspond to the bin size.  The gray symbols indicate the results for the ensemble of the mocks with the error bars indicating the 16th--84th percentiles.  The green circles corresponds to the contaminants of the mocks.  The corresponding thick lines indicate the averages in steps of $2\arcmin$.}
   \label{fig:radialcount}
\end{figure}

We now revisit our LBG candidate sample in the J0148$+$0600 field and compare the observation with the mock, taking into account the spatial effects.  In Figure \ref{fig:radialcount}, we present the surface density of LBG candidates as a function of radius from the quasar position.  Masked areas are accounted for to calculate the surface densities by using the random-point catalog (see Section \ref{sec:data}).  The upper panel of Figure \ref{fig:radialcount} shows the cumulative surface densities within a given radius in steps of $1\arcmin$ out to $40\arcmin$.  The horizontal line indicates the mean surface density (0.041~arcmin$^{-2}$).  In order to compare with a uniform distribution including the effects of spatially varying photometric noise, we computed the radial profile by using the mock catalogs simulated using the UD-COSMOS data.  In the figure, we show the radial profile for the mock samples of all candidates (gray) and possible contaminants (green).  The hatched areas indicate the 16th--84th percentiles of the the 500 total realizations.  Note that the mock catalogs predict a slightly higher mean surface density ($\left< \bar{\Sigma}^\mathrm{mock}_\mathrm{LBGc} \right> = 0.053~\mathrm{arcmin}^{-2}$) including contamination.  

Likewise, in the lower panel of Figure \ref{fig:radialcount}, we show the surface densities computed in radial bins of width $8\arcmin$, or $\approx 19~\mathrm{cMpc}$ (dashed concentric circles in Figure \ref{fig:LBG_maps}).  The horizontal error bars correspond to the size of the radial bins.
For the mock, we show the ensemble average in the radial bins and the 16th--84th percentiles by error bars for all the candidates (gray) and the contaminants (green).  In addition, we plot the ensemble averages computed in $2\arcmin$ width bins.

In both panels of Figure \ref{fig:radialcount}, the mock catalogs show a moderate increase in the expected surface density with increasing radius, reflecting the radially increasing noise levels.  This radial trend arises due to the fact that the expected number density of contamination increases with radius.  However, comparisons between the data and the mock clearly indicate that the actual LBG candidate sample presents a much stronger decrement toward the field center than that arising due to the spatially varying noise.

As a further test, we examine a possible brightness dependence of the radial trend.  We separated LBG candidates equally into two subsample according to the observed $z_\mathrm{AB}$ magnitude at $z_\mathrm{AB}=25.08$ and computed the surface densities in the same way as above.  The results are shown with open circles in Figure \ref{fig:radialcount}.  The radial trends of these brighter and fainter subsamples are in excellent agreement.  This suggests that there is a real underdensity of galaxies near the quasar position.

\begin{figure}[tbp] 
   \centering
      \includegraphics[width=3.5in]{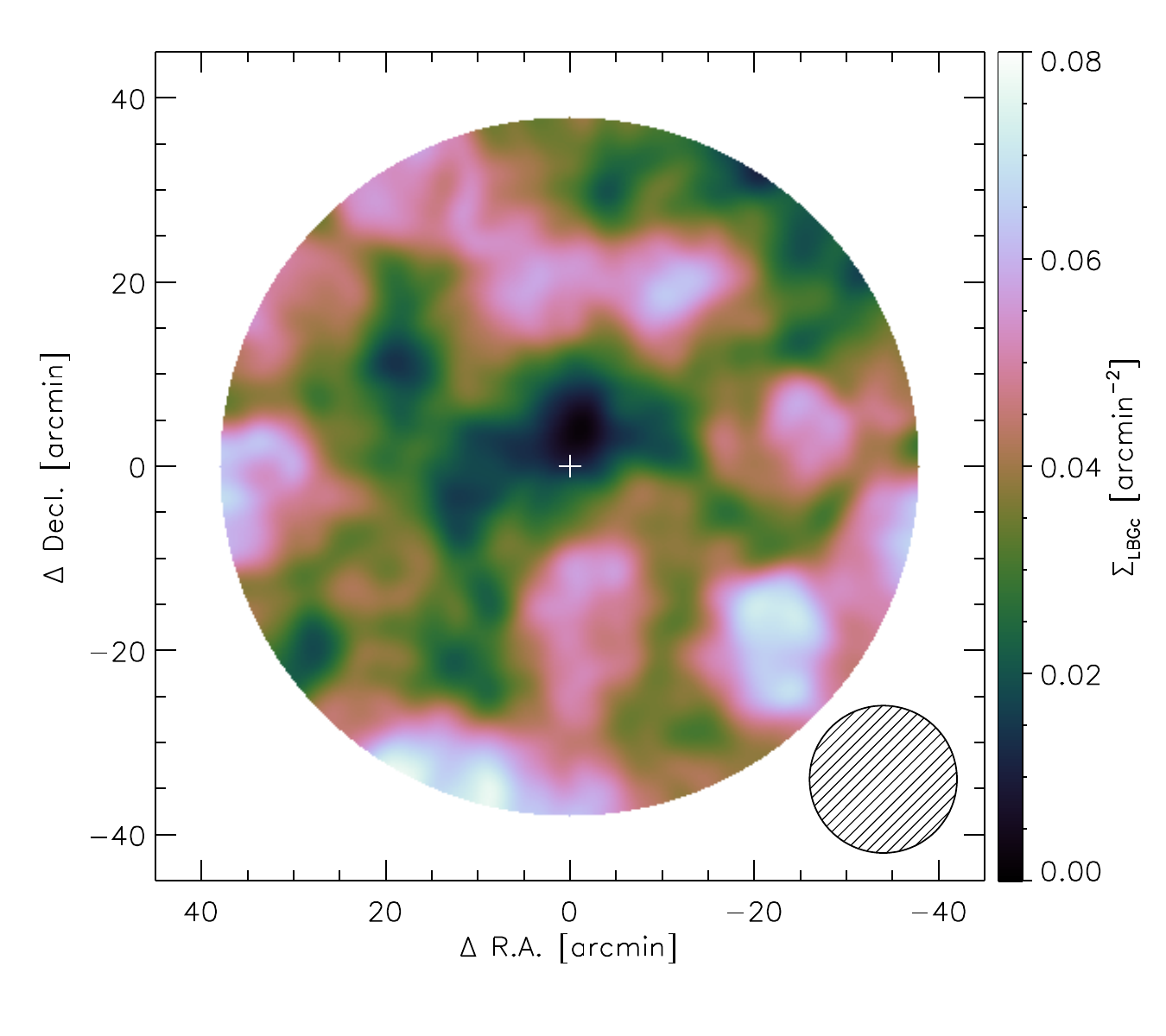}
   \caption{Left: surface density map of LBG candidates shown with color gradation.   The surface density at a given position is calculated from the number of LBG candidates within a fixed aperture of $8\arcmin$ in radius and corrected for the spatial effects due to the spatially varying noise.  The surface density map is truncated within $38\arcmin$ of the field center.}
   \label{fig:LBG_densitymap}
\end{figure}

To gain further information on the environment of the Ly$\alpha$ trough of J0148+0600, we created the surface density map as follows.  We counted LBGs in circular apertures centered on each point of a regular grid of $0.\!\arcmin2$ spacing.  Each number count was converted into a surface density accounting for masked areas and the fraction of areas outside the survey area.  Raw surface densities were then corrected for the spatial effects by a factor $\left<\bar{\Sigma}^\mathrm{mock}_\mathrm{LBGc}\right>/\left< \Sigma^\mathrm{mock}_\mathrm{LBGc}(\alpha,\delta) \right>$ where $\left< \Sigma^\mathrm{mock}_\mathrm{LBGc}(\alpha,\delta) \right>$ is the map shown in Figure \ref{fig:map_predicted} and the numerator is the mean across the field.  The surface density map is smoothed using a Gaussian kernel of $\sigma=1.\!\arcmin2$ (or $\approx 2.8~\mathrm{cMpc}$) and truncated within $38\arcmin$ of the field center.  In Figure \ref{fig:LBG_densitymap}, we show the results.  The lowest surface density is measured near the quasar position ($\Delta \mathrm{R.A.}=-1.\!\arcmin4$; $\Delta \mathrm{decl.}=+4.\!\arcmin0$).  

\subsection{Significance estimates}

\begin{figure}[tbp] 
   \centering
      \includegraphics[width=3.5in]{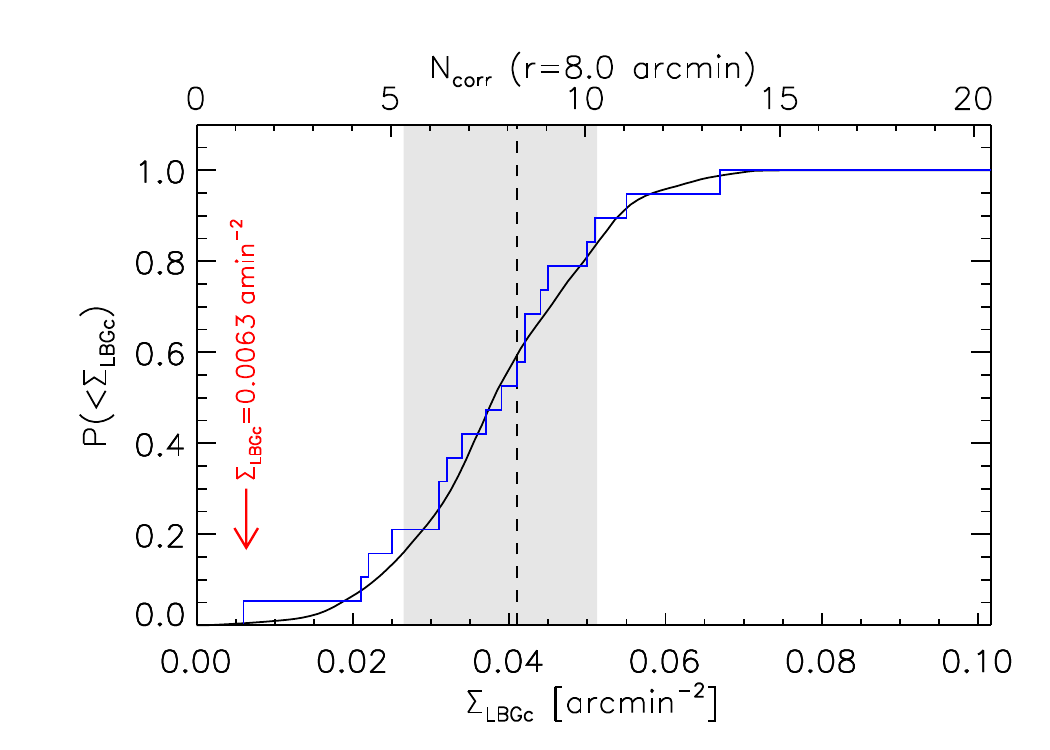}
   \caption{Cumulative distribution of number counts at each grid points.  The number counts are corrected for the effects of the inhomogeneous noise levels.  Note that the individual measurements are highly correlated with the measurements in their vicinity.  The step-like function corresponds to 19 independent apertures which are not overlapping with each other.  The vertical red line indicates the value at the quasar position.  The vertical dashed line and the shaded region indicate the mean value and the central 68th percentiles.  The top $x$-axis translates angular scales to comoving distances at $z=5.7$.}
   \label{fig:densityhist}
\end{figure}

We here evaluate the significance of the LBG underdensity near the quasar J0148$+$0600.  In Figure \ref{fig:densityhist}, we show the cumulative probability distribution of the surface densities.  The surface density at the quasar position is $0.0063~\mathrm{arcmin}^{-2}$ after the correction for the spatial effects, which is marked by an arrow.  The step-like function indicates the measurements within 19 independent circular apertures that are not overlapping in the survey field.  One of these apertures is centered on the quasar position, where the lowest surface density is measured. In addition, the smooth curve corresponds to the surface density measurements at all $0.\!\arcmin2$ spaced grid points within $38\arcmin$.  The shaded region corresponds to the 16th--84th percentiles, which are slightly biased toward lower values relative to the mean surface density due to a skewed distribution of $\Sigma_\mathrm{LBGc}(\alpha,\delta)$.  The fraction of grid points where the surface densities are lower than the value centered on the quasar is found to be only 0.009 and that almost all (96\%) of these grid points are within $8\arcmin$ of the center.  This indicates nearly zero ($<0.1\%$) probability measuring a surface density equal to or lower than the value at the field center when taking a circular aperture at a random position at $R>8\arcmin$.

It is worth commenting on other smoothing scales.  The central aperture of $r=6\arcmin$ ($10\arcmin$) counts zero (five) LBG candidates, which remains the single lowest value among 35 (12) independent circular regions.  We can therefore be confident of the presence of the underdensity near the field center.  Moreover, we would emphasize that the center of the hole, where we measured the lowest surface density, is quite close to the quasar position but is somewhat displaced as might be expected.  Therefore, the association between the exceptional Ly$\alpha$ trough of J0148$+$0600 and a region that is highly underdense in bright LBGs appears robust. 

Lastly, we give a naive assessment in terms of Poisson statistics by using the mock catalogs.  It is must be kept in mind, however, that the distribution of LBGs in the real universe is not a random process and thus, strictly speaking, the application of a Poisson probability distribution is not suitable.  We compare the count of the candidates ($N_\mathrm{LBGc}=1$) within $\le 8\arcmin$ of the quasar position with the mean count (6.8) that is expected including the spatial effect from the observed background surface density ($\Sigma_\mathrm{LBGc}(R>15\arcmin)=0.044~\mathrm{arcmin^{-2}}$) that is measured outside $15\arcmin$ of the quasar position.  If a Poisson distribution with this mean value is assumed, the probability of detecting $\le 1$ candidate is only 0.009.  The presence of the void is also supported by the fact that, among the 500 mocks, there is no realization with such a small number count ($N_\mathrm{LBGc}(<8\arcmin)=1$) in the same region (minimum $N^\mathrm{mock}_\mathrm{LBGc}(<8\arcmin)=2$).

\section{Discussions} \label{sec:discussions}
\subsection{Comparisons to models}

The reason why the extreme Ly$\alpha$ trough of J0148$+$0600 draws particular attention is that the difference in the predicted galaxy surface density between the fluctuating-$\lambda_\mathrm{mfp}$ model \citep{2016MNRAS.460.1328D} and the fluctuating-$T_\mathrm{IGM}$ model \citep{2015ApJ...813L..38D} becomes unambiguously large at $\taueff\sim7$ , as demonstrated in semianalytic simulations by \citet{2018ApJ...860..155D}.  

In the fluctuating-$\lambda_\mathrm{mfp}$ model, opaque troughs arise from voids where the ionizing background is suppressed and subsequently the hydrogen neutral fraction is relatively high.  The seminumerical simulations predict an average surface density $\lesssim 0.4$ of the mean density within $\sim 10~\mathrm{cMpc}$ with large scatter.

In the fluctuating-$T_\mathrm{IGM}$ model, the spatial variations in the IGM temperature arise due to variations in the timing of reionization.  Higher opacities occur in regions that reionized earlier because such regions have had longer time to cool, increasing the neutral fraction with a relatively high recombination rate at the epoch of interest, i.e, $z\sim5.7$.  In the excursion-set formalism of reionization \citep{2004ApJ...613....1F}, it is expected that overdense regions reionize in earlier epochs, and thus the most opaque troughs would arise from the highest overdensities.

\begin{figure}[tbp] 
   \centering
      \includegraphics[width=3.5in]{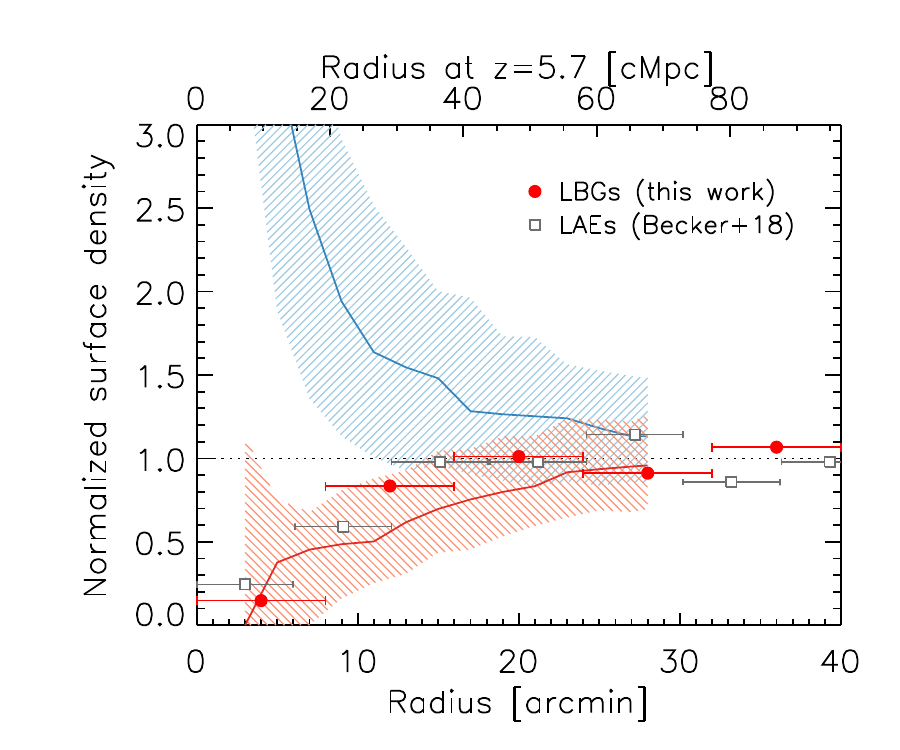}
   \caption{Radial surface density normalized by the background values.  Red circles are the same as in Figure \ref{fig:radialcount}, but corrected for the effects due to radially increasing noise and normalized by $\Sigma_\mathrm{LBGc}(R>15\arcmin)=0.044~\mathrm{arcmin^{-2}}$.  The red and blue lines indicate the predicted median radial profile for 160~cMpc sightlines with $\taueff>6.5$ in the fluctuating-$\lambda_\mathrm{mfp}$ model and the fluctuating-$T_\mathrm{IGM}$ model, respectively, taken from \citet[][Figure~10]{2018ApJ...860..155D}.  The hatched region corresponds to the central 68th percentiles of the galaxy counts in the simulations.  The observed relative density trend is in good agreement with the prediction for the fluctuating-$\lambda_\mathrm{mfp}$ model.  The top $x$-axis translates angular scales to comoving distances at $z=5.7$.  For reference, the squares show the measurements of \citet{2018ApJ...863...92B} for LAEs at $z\approx5.7$ in the same field.}
   \label{fig:comp_with_Davies18}
\end{figure}

We attempt to compare the observed radial surface density profile with the predictions for these models.  We count LBG candidates within five annular bins of $8\arcmin$ (see Figure \ref{fig:radialcount}) and correct these raw surface density values for the effects due to radially increasing noise by a factor $\left<\bar{\Sigma}^\mathrm{mock}_\mathrm{LBGc}\right>/\left<\Sigma^\mathrm{mock}_\mathrm{LBGc}(R)\right>$, where the denominator is the mean number density within the radial bins of the mocks (see Section \ref{sec:mock}).  These corrections range from $+15\%$ at the innermost bin to $-5\%$ of the outermost bin.  The surface densities are then normalized by the background surface density ($\Sigma_\mathrm{LBGc}(R>15\arcmin)=0.044~\mathrm{arcmin^{-2}}$) that is measured outside $15\arcmin$ of the quasar position.  The predicted surface density profiles along 160~cMpc sightlines with $\taueff>6.5$ are taken from Figure 10 of \citet{2018ApJ...860..155D} and normalized by the mean number density.  

Before comparison, it should be noted that the predicted profiles are computed for simulated galaxies brighter than $M_\mathrm{UV}=-21$, or apparent magnitude $m_\mathrm{UV}\lesssim 25.7$, whereas the apparent magnitude range of our LBG sample is limited to $24\le z_\mathrm{AB}\le 25.3$.  In addition, the predictions are computed from galaxy counts using a top-hat selection function along the simulated sightlines though the actual selection function is smoothed, as shown in Figure \ref{fig:completeness}.  We do not regard these differences as significant for distinguishing these two models.

Figure \ref{fig:comp_with_Davies18} compares the data to the predictions from the \citet{2018ApJ...860..155D} simulations.  We find that the observed radial trend of the normalized surface density is in good agreement with the prediction for the fluctuating-$\lambda_\mathrm{mfp}$ model.  In contrast, the data strongly disfavor the fluctuating-$T_\mathrm{IGM}$ model that predicts a striking increase in galaxy surface density within $15\arcmin$ of the sightline.  These conclusions are consistent with the statement made by \citet{2018ApJ...863...92B}.

We, however, need to be careful before settling the question.  These two scenarios assume that the universe reionized completely by a redshift of $z\ge6$, higher than the epoch of observations.  Therefore, the predicted correlations between $\taueff$ and density are for the ``post-reionization'' IGM.  In particular, the simulations of the fluctuating-$T_\mathrm{IGM}$ scenario impose that the simulation box is reionized in its entirety by $z=6$, and thus all regions experience the photoionization heating at $z\ge6$.  However, if there remain some regions that were not reionized at $z\sim5.7$, such regions, which should be lowest densities, do not take part of the post-reionization relationship between $\taueff$ and density.  This could happen within the uncertainties of the Thomson scattering optical depth \citep{2016A&A...594A..13P}.  If the Ly$\alpha$ trough of the J0148$+$0600 field arises from such a rare void that it was not yet reionized, then the conclusion above would not be justified.  It would be fair, however, to treat such scenarios with substantial regions not reionized at $z<6$ as separate from the fluctuating-$\lambda_\mathrm{mfp}$ and fluctuating-$T_\mathrm{IGM}$ models.

The possible presence of underdense ``neutral islands'' at $z<6$ has been indeed considered in the late-reionization model \citep{2019MNRAS.485L..24K,2020MNRAS.491.1736K}, which our results are also consistent with.  \citet{2020MNRAS.491.1736K} show in their Figure 7 that the surface density of bright LBGs begins to decline at a relatively large radius of $\sim70~\mathrm{cMpc}$.  However, in this model, underdense regions at the same level as the one where the remarkable trough is seen could also present a low $\taueff\sim2$, because such regions may reionize shortly before and thus the neutral islands vanish.  This is a critical difference from the fluctuating-$\lambda_\mathrm{mpf}$ model, in which low-$\taueff$ regions should always be associated with overdensities.  Therefore, further observations, especially, to characterize the density variation in low-$\taueff$ regions are important to distinguish between the most plausible models (i.e., the fluctuating-$\lambda_\mathrm{mfp}$ and late-reionization models) to date.

\subsection{Comparison to the LAE map}

We here compare the void of LBG candidates to that discovered in LAEs at $z\approx5.72$.  Cross-matching our catalog to the \citet{2018ApJ...863...92B} LAE sample, we found that only $\approx3\%$ of the LAEs have $z_\mathrm{AB}\le 25.3$, i.e., the limiting magnitude for the LBG selection, and that the stacked mean flux of LAEs is $\left< z_\mathrm{AB}\right>=26.8$.  This indicates that the LAE sample would trace less massive galaxies, which are more dominant in the total number density of galaxies, while our LBG sample traces more massive galaxies, which would be more biased relative to the matter distribution.  Therefore, although our result alone cannot rule out the presence of fainter LBGs in the vicinity of the quasar sightline, the combination of these two suggests that, at least in a narrow redshift range at $z\approx5.72\pm0.05$, the paucity of galaxies holds true across a range of stellar masses.

In Figure \ref{fig:comp_with_Davies18}, we overplot the LAE surface density profile taken from \citet{2018ApJ...863...92B}.  The radial profiles of the LBG and LAE samples are in agreement.  Both of these results indicate that the transverse diameter of the underdense region giving rise to the trough is $\sim 60~\mathrm{cMpc}$.  Because the redshift window for LBGs is as wide as the length of the trough ($\sim160~\mathrm{cMpc}$), the results suggest the presence of an elongated region that is coherently underdense.  This may be a single prolate void, or consist of multiple void regions which overlap in projection.

Given the relatively poor statistics, especially, of our LBG sample and its relaxed redshift distribution, more detailed comparisons of the density structures are challenging.  Such comparisons, however, have the potential to provide further constraints on the reionization scenario.  The presence of neutral islands, as predicted in the late-reionization model, could be examined by investigating the suppression of Ly$\alpha$ that might result in a void relatively deeper than that of LBGs.  Similarly, the spatial variation in the typical EW of Ly$\alpha$ would provide insights in this context.  This issue will be addressed with upcoming data from ongoing HSC observations in a future paper.

\subsection{Bright quasar candidates}

Lastly, we comment on possible high-$z$ quasars in the J0148$+$0600 field.   In the rare-source model of \citet{2017MNRAS.465.3429C}, the observed large $\taueff$ fluctuations and spatial coherence arise due to fluctuations in a UVB with significant contribution from rare bright sources.  The presence or absence of these rare sources fluctuates the local UVB.  This model predicts that we should always find moderately luminous quasars ($M_\mathrm{UV}<-22$, equivalent to $m_\mathrm{UV}\lesssim 24.7$) within $\sim50~\mathrm{cMpc}$ of low-$\taueff$ regions, but not near high-$\taueff$ regions.  

\begin{figure}[tbp] 
   \centering
      \includegraphics[width=3.5in]{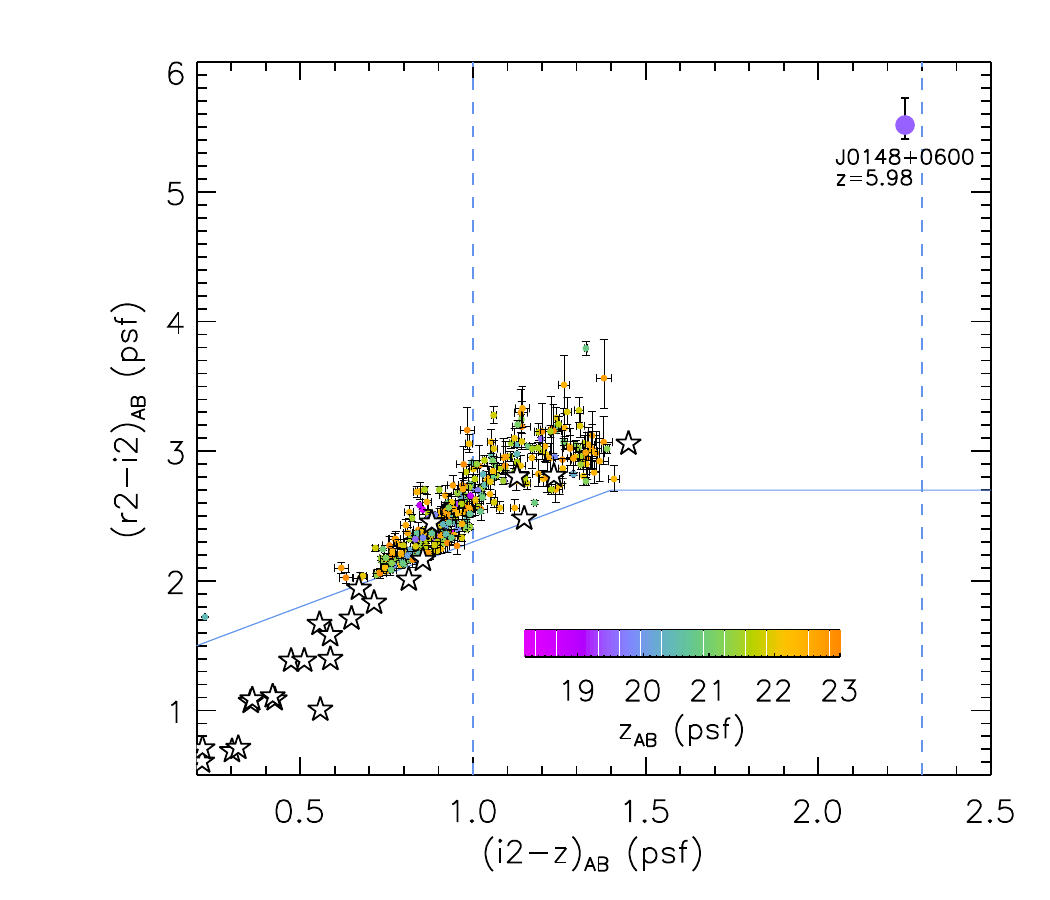}
   \caption{The ($i2-z$) versus ($r2-i2$) color diagram for unextended sources using forced PSF photometry.  Small circles indicate those brighter than  $z_\mathrm{AB}\le 23$ and lying above the ($r2-i2$) threshold for the LBG selection (blue solid line).  The large circle at the upper right indicates the central quasar J0148$+$0600 at $z=5.98$.  These symbols are color-coded by their $z$ magnitude according to the color scale in the panel.  Vertical dashed lines indicate the ($i2-z$) limits for LBG selection.  The white stars indicate Galactic M stars from \citet{1983ApJS...52..121G}.}
   \label{fig:iz_vs_ri_qso_psf}
\end{figure}

Our wide-field HSC data allow us to try to test the rare-source model by searching for quasar candidates in the regime brighter than LBGs.  We assume that such objects are to be detected as point sources.  In Figure \ref{fig:iz_vs_ri_qso_psf}, we show the ($i2-z$) versus ($r2-i2$) color diagram for point-like sources using forced PSF photometry.  Sources shown here are limited to be brighter than $z_\mathrm{AB}\le 23$ and to lie above the ($r2-i2$) threshold for the LBG selection.  The central quasar J0148$+$0600 at $z=5.98$ itself is marked by the large circle at the upper right.  These symbols are color-coded by their $z$ magnitude according to the inset color scale.  Though the model considers quasars down to $M_\mathrm{UV}<-22$, the magnitude range we explored here is limited to $z_\mathrm{AB} \le 23$ (i.e., $M_\mathrm{UV}\lesssim -23.6$, assuming $z_\mathrm{AB} \approx m_\mathrm{UV}$) because the ($r2-i2$) color is poorly constrained for fainter objects and thus no reasonable classification can be made.

Figure \ref{fig:iz_vs_ri_qso_psf} shows that there are no obvious quasars with a ($r2-i2$) color as red as J0148$+$0600 and that all the possible candidates are in fact located near the locus of known Galactic M stars.  The currently available ($r2$, $i2$, $z$) colors do not allow for robust classification between high-$z$ quasars and stars.  Therefore, we argue that the absence of unambiguously luminous quasars is certainly consistent with the expectation from the rare-source model, but cannot rule out the presence of quasars with colors similar to stars and/or less bright quasars which dominate the total number density of ionizing sources in this model.  Examining the absence of such quasars in high-$\taueff$ regions requires spectroscopic confirmation to distinguish them from stars.  On the other hand, the discovery of bright quasars in low-$\taueff$ fields can immediately be a supporting indication of this scenario.  Observations in low-$\taueff$ fields will thus be essential to test the role of rare quasars as a possible origin of the observed large $\taueff$ fluctuations.  

\section{Summary} \label{sec:summary}

We have carried out a search for LBGs at $5.5\lesssim z\lesssim 5.9$ in the field of the $z=5.98$ quasar J0148$+$0600, whose spectrum shows an extremely long ($\sim 160~\mathrm{cMpc}$) and opaque ($\taueff \gtrsim 7$) Ly$\alpha$ trough across the relevant redshift range.  Our goal was to test different models proposed to explain the observed large $\taueff$ fluctuations at $z\sim5.5\textrm{--}6.0$.  In the fluctuating-$\lambda_\mathrm{mpf}$ model, the $\taueff$ fluctuations are due to fluctuations in the galaxy-dominated UVB with spatially varying mean free path, and, consequently, opaque troughs such as those seen in the J0148$+$0600 quasar spectrum could arise in a low-density region \citep{2016MNRAS.460.1328D}.  The late-reionization model predicts a similar situation, but due to residual neutral islands instead of a relatively high neutral fraction for a weak UVB \citep{2019MNRAS.485L..24K,2020MNRAS.491.1736K}.  The rare-source model attributes possible enhancement in the UVB fluctuations to significant contributions of rare, bright sources such as quasars, and predict that we will find no such source within $\sim100~\mathrm{cMpc}$ of the trough \citep{2015MNRAS.453.2943C,2017MNRAS.465.3429C}.  The fluctuating-$T_\mathrm{IGM}$ model explains the $\taueff$ distribution with relic fluctuations in the gas temperature and predicts that such troughs should arise in high-density regions which were ionized earlier \citep{2015ApJ...813L..38D}.

We constructed a sample of 185 LBG candidates within $40\arcmin$ of the quasar position down to $z_\mathrm{AB}=25.3$ in the $2.\!\arcsec0$ diameter aperture photometry.  The color criteria were optimized so that the expected redshift distribution has a peak at the center of the trough and covers its entire redshift range.  The spatial distribution of the LBG candidates shows a pronounced paucity within $8\arcmin$, or $\approx19~\mathrm{cMpc}$ of the quasar position.  We compare the observed spatial distribution of the candidates to those expected for an intrinsically uniform distribution with spatially varying photometric noise and find that the observed underdensity cannot be ascribed to the instrumental spatial effects.  The association between this extraordinary Ly$\alpha$ trough and the LBG underdensity is ensured by the fact that the probability of measuring a surface density equivalent to or lower than that at the quasar position is only $<1\%$ when taking random positions across the field.

Our results are consistent with the results from an LAE survey of \citet{2018ApJ...863...92B}, and suggest that the \citet{2018ApJ...863...92B} result is not purely due to suppression of Ly$\alpha$ emission by the high opacity, but reflects a true paucity of galaxies.  Furthermore, our LBG candidate sample covers the full length of the trough ($\sim 160~\mathrm{cMpc}$), instead of a narrow redshift range probed by the LAEs in a narrowband filter, and thus suggests that there is a spatially coherent underdensity associated with the trough.

The association of a real underdensity of galaxies and a region with extremely high $\tau_\mathrm{eff}$ is consistent with what is predicted by the fluctuating-$\lambda_\mathrm{mfp}$ model and the late-reionization model.  The absence of obviously luminous quasars around the trough is also consistent with the rare-source model.  In contrast, the fluctuating-$\lambda_\mathrm{IGM}$ model appears to be strongly disfavored because it predicts that the trough should be associated with an overdensity (see Figure \ref{fig:comp_with_Davies18}).  For more conclusive statements, we require further observations in multiple quasar fields, including those with both extremely high and low $\taueff$.  We will present in future papers comprehensive studies using the full set of our HSC observations.

\acknowledgements

This work has been supported by the Swiss National Science Foundation.

We are grateful to the Subaru telescope staff, especially the HSC queue team and the pipeline team for their expertise.  
This paper is based on data collected at the Subaru Telescope and retrieved from the HSC data archive system, which is operated by Subaru Telescope and Astronomy Data Center (ADC) at National Astronomical Observatory of Japan (NAOJ).  Data analysis was carried out on the Multi-wavelength Data Analysis System operated by ADC, NAOJ.

The Hyper Suprime-Cam (HSC) collaboration includes the astronomical communities of Japan and Taiwan, and Princeton University. The HSC instrumentation and software were developed by NAOJ, the Kavli Institute for the Physics and Mathematics of the Universe (Kavli IPMU), the University of Tokyo, the High Energy Accelerator Research Organization (KEK), the Academia Sinica Institute for Astronomy and Astrophysics in Taiwan (ASIAA), and Princeton University. Funding was contributed by the FIRST program from Japanese Cabinet Office, the Ministry of Education, Culture, Sports, Science and Technology (MEXT), the Japan Society for the Promotion of Science (JSPS), Japan Science and Technology Agency (JST), the Toray Science Foundation, NAOJ, Kavli IPMU, KEK, ASIAA, and Princeton University. 

This paper makes use of software developed for the Large Synoptic Survey Telescope. We thank the LSST Project for making their code available as free software at \url{http://dm.lsst.org}.  

The Pan-STARRS1 Surveys (PS1) have been made possible through contributions of the Institute for Astronomy, the University of Hawaii, the Pan-STARRS Project Office, the Max-Planck Society and its participating institutes, the Max Planck Institute for Astronomy, Heidelberg and the Max Planck Institute for Extraterrestrial Physics, Garching, The Johns Hopkins University, Durham University, the University of Edinburgh, Queen’s University Belfast, the Harvard-Smithsonian Center for Astrophysics, the Las Cumbres Observatory Global Telescope Network Incorporated, the National Central University of Taiwan, the Space Telescope Science Institute, the National Aeronautics and Space Administration under Grant No. NNX08AR22G issued through the Planetary Science Division of the NASA Science Mission Directorate, the National Science Foundation under Grant No. AST-1238877, the University of Maryland, and Eotvos Lorand University (ELTE) and the Los Alamos National Laboratory.

This work has made use of data from the European Space Agency (ESA) mission
{\it Gaia} (\url{https://www.cosmos.esa.int/gaia}), processed by the {\it Gaia}
Data Processing and Analysis Consortium (DPAC,
\url{https://www.cosmos.esa.int/web/gaia/dpac/consortium}). Funding for the DPAC
has been provided by national institutions, in particular the institutions
participating in the {\it Gaia} Multilateral Agreement.

{\it Facility}: Subaru (HSC)

\bibliographystyle{apj}
\bibliography{apj-jour,ads}

\end{document}